\documentclass[
 reprint,
 amsmath,amssymb,
 aps,
]{revtex4-2}

\usepackage{graphicx}
\usepackage{dcolumn}
\usepackage{bm}
\usepackage{color}

\usepackage{soul,xcolor}
\usepackage[normalem]{ulem} 

\begin{document}
\setstcolor{red}
\preprint{APS/123-QED}
\title{Disorder-Induced Anomalous Mobility Enhancement in Confined Geometries
}

\author{Dan Shafir}
\email{dansh5d@gmail.com}
\affiliation{Physics Department, Bar-Ilan University, Ramat Gan 5290002,
Israel}
\author{Stanislav Burov}
\email{stasbur@gmail.com}
\affiliation{Physics Department, Bar-Ilan University, Ramat Gan 5290002,
Israel}


\begin{abstract}

Strong, scale-free disorder disrupts typical transport properties like the Stokes-Einstein relation and linear response, leading to anomalous, non-diffusive motion observed in amorphous materials, glasses, living cells, and other systems. 
 Our study reveals that the combination of scale-free quenched disorder and geometrical constraints induces unconventional single particle mobility behavior.
  Specifically, in a $2$-dimensional channel with width $w$, 
 under external drive, tighter geometrical constraints (smaller 
$w$) enhance mobility. 
We derive an explicit form of the response to an external force by utilizing the double-subordination approach for the quenched trap model. The observed mobility enhancement occurs in the low-temperature regime where the distribution of localization times is scale-free.

\end{abstract}

\maketitle

\newcommand{\avg}{\langle \tau_i \rangle}
\newcommand{\fr}{\frac}
\newcommand{\tl}{\tilde}

Transport in disordered and amorphous materials has attracted vast attention for many decades \cite{scott2023extracting, waigh2023heterogeneous, bouchaud1990anomalous, metzler2000random, montroll1979enriched, metzler2014anomalous, barkai2012single, anderson2019filament, yamamoto2021universal, metzler2022modelling}.
The study of systems' response to external forces, particularly with an aim to optimize transport, constitutes an imperative focus of research~\cite{malgaretti2019special}.
The external force can be a result of an electric field pulling on an electron through a conductor \cite{kondrat2014accelerating, kondrat2014charging, benichou2018tracer} or a pressure gradient pushing on a molecule diffusing in a channel \cite{uspal2013engineering, sarracino2016nonlinear, cecconi2017anomalous, cecconi2018anomalous}. 
The classical depiction of such dynamics is Drude’s model of current flow in a metal~\cite{ashcroft1976solid}. It describes, through the application of kinetic theory, the diffusion of electrons by repeated encounters with immobile hard scatterers (such as ions or impurities).
When an external electric field is applied, the charge carriers experience a net drift velocity related to the mean free path between scattering events. 
The resulting picture is that the response to the external force, i.e., carrier mobility, is an intrinsic property of the medium. Therefore, for transport in restricted geometry, like a channel, the expectation is that the mobility will be independent of channel width (or cross-section) or sometimes will increase with channel width due to the availability of new pathways.
In this work, we explore the mobility properties of transport inside a channel with the presence of strong and quenched disorder. 
Specifically, we aim to demonstrate that quenched and strong disorder can redefine our understanding of the dependence of mobility on geometry. 

Experiments in amorphous materials~\cite{scharfe1970transient, pai1972charge, pfister1974pressure, mort1970steady, gill1972drift} have shown that a packet of charge carries do not propagate in a Gaussian manner and instead exhibit a dispersion of carrier transit times. Scher and Montroll~\cite{scher1975anomalous, montroll1979enriched} termed this phenomenon anomalous transport and suggested that carriers are affected by deep traps or local areas of arrest. When the duration $\tau$ of such events follows a power-law distribution, i.e. $\sim \tau^{-1-\alpha}$, and $0<\alpha<1$, the transport becomes subdiffusive~\cite{bouchaud1990anomalous}.
Meaning that the mean squared displacement (MSD) is not proportional to time $t$ but rather grows sublinearly, i.e. MSD$\sim t^\alpha$, as observed in amorphous materials~\cite{scharfe1970transient, pai1972charge, gill1972drift, pfister1974pressure, mort1970steady, yuan2000time, tyutnev2017mobility, dunlap1996hopping, scher1975anomalous}, biological cells~\cite{scholz2016cycling, metzler2014anomalous, barkai2012strange, kompella2024determines, yu2018subdiffusion, anderson2019filament, sabri2020elucidating}, granular materials~\cite{marty2005subdiffusion, bodrova2024diffusion, metzler2022modelling}, non-Newtonian fluids~\cite{wong2004anomalous} and other systems~\cite{meyer2024time, mckinley2009transient, paoluzzi2024flocking}.
These power-law distributed waiting times ($\sim \tau^{-(1+\alpha)}$), as detected in various systems~\cite{tabei2013intracellular,wong2004anomalous,vilk2022ergodicity, vilk2022classification, yamamoto2021universal}, can appear naturally due to the exponential distribution of the depths of energetic wells that give rise to the regions of local arrest.
The strong disorder ($0<\alpha<1$) results in a diverging mean trapping time that disrupts regular diffusive properties and leads to aging, weak ergodicity breaking, and non-self averaging~\cite{bouchaud1992weak, monthus1996models, rinn2000multiple, rinn2001hopping, bertin2003subdiffusion, burov2007occupation}. 
Most theoretical studies address the annealed version of the strong disorder. Namely, the waiting times in the trapping regions are uncorrelated, and each time the particle returns to the same arrest region, it is localized for a different time. 
Such framework was termed the continuous time random walk (CTRW)~\cite{scher1975anomalous, bouchaud1990anomalous, metzler2014anomalous, hamdi2024laplace}, a very popular model of anomalous transport. 
The quenched version with a strong disorder, termed the quenched trap model (QTM), treats the trapping times during revisits of the arrest region as correlated.
For the QTM regular techniques and Stokes-Einstein–Smoluchowski theory do not apply due to strong correlations and memory effects \cite{bouchaud1990anomalous, akimoto2018non, akimoto2020trace}.
Scaling arguments and renormalization group approach~\cite{bouchaud1990anomalous, machta1985random, monthus2003anomalous} among other works~\cite{ben2006aging, arous2007scaling, ben2015randomly, vcerny2015randomly, burov2017quenched} suggest that for dimensions $d >2$, QTM behaves  qualitatively as CTRW in the subdiffusion regime. But big differences can be witnessed as we show.

In this work we explore the effects of a strong quenched disorder on particle mobility under the geometrical constraint of a channel with width $w$.
By utilizing the recently developed double-subordination technique~\cite{shafir2022case, burov2020transient, burov2012weak, burov2017quenched, burov2011time}, we obtain an analytical expression for the mobility and its dependence on the external driving force $f$, temperature $T$ and width $w$. 
For low temperatures,
we find that the mobility is a decreasing function of $w$. 
Namely, the response to an external drive weakens as the channel cross-section grows. 
Such a counterintuitive enhancement with decreasing $w$ appears only when the disorder is quenched and strong. When one of these requirements is omitted, the mobility is independent of the channel width. 

\textit {The quenched trap model.}
The physical picture behind QTM is a thermally activated particle jumping between energetic traps.
When a particle is in a trap, the average escape time $\tau$ is provided by the Arrhenius law $\tau \propto \exp \left(E_{\mathbf{r}} / T\right)$, where $E_{\mathbf{r}}>0$  is the depth of trap at position $\mathbf{r}$
and $T$ is the temperature. 
When the distribution of the energetic traps $E_{\mathbf{r}}$ is exponential, $\phi(E_{\mathbf{r}})=\exp (-E_{\mathbf{r}} / T_g) / T_g$, the distribution of the average escape time is 
\begin{eqnarray} \label{eq: waiting time PDF}
    \psi\left(\tau_{\mathbf{r}}\right) \sim \tau_{\mathbf{r}}^{-1-\frac{T}{T_g}} A /|\Gamma(-T/T_g)|,
\end{eqnarray} 
$A=|\Gamma(-T/T_g)| T/T_g$ and $\Gamma(\dots)$ is the Gamma function. For $T<T_g$, the slow power-law decay of $\psi(\tau)$ leads to a diverging mean escape, when averaged over disorder.
In the following, we set $\alpha=T/T_g$ and focus on the glassy regime $0<\alpha<1$, where  QTM exhibits aging and non-self-averaging behavior~\cite{akimoto2018non, akimoto2020trace, monthus2003anomalous}.
The average escape times $\tau_{\mathbf{r}}$ serve in QTM as the waiting times. Each time the particle visits position $\mathbf{r}$, it spends there exactly the same time $\tau_{\mathbf{r}}$ hence the disorder is quenched. 
In~\cite{bertin2003subdiffusion}, no difference was found between setting quenched waiting times or setting quenched average waiting times.
The quenched variables $\{ \tau_{\mathbf{r} }\}$ are positive, independent, identically distributed (i.i.d) random variables with probability density function (PDF) provided by Eq.~(\ref{eq: waiting time PDF}).
We consider the spatial process between different positions $\mathbf{r}$ as a random hop process on a two-dimensional square lattice with lattice spacing $a$ taken to be $1$ (a.u).
At time $t = 0$, the particle starts at $\mathbf{r} = 0$ and stays at this position for the period $\tau_{\mathbf{0}}$ before jumping to some random site ${\mathbf{r}}^{\prime}$ where it waits for the period $\tau_{\mathbf{r}^{\prime}}$ and then the random jump + waiting period procedure continues.
The probability of transition (jump) from $\mathbf{r}$ to $\mathbf{r^{\prime}}$ is provided by $p(\mathbf{r}^{\prime};\mathbf{r})$.
We assume that the lattice is translationally invariant in space, i.e.,  $p(\mathbf{r}^{\prime};\mathbf{r})$ is a function of $\mathbf{r}^{\prime}-\mathbf{r}$: $p(\mathbf{r}^{\prime} - \mathbf{r})$.
The disorder averaged positional PDF of finding the particle at position $\mathbf{r}$ at time $t$, $\langle P(\mathbf{r}, t)\rangle$ ($\langle \cdots \rangle$ represents the averaging over disorder) is found by utilizing the double subordination technique~\cite{shafir2022case, burov2020transient} that we briefly describe below.

\textit{The diffusion front}.
The effect of correlations imposed by quenched disorder is appreciated when the measurement time $t$ is written in terms of the local waiting times $\tau_{\mathbf{r}}$. Namely, $t=\sum_{\mathbf{r}} n_{\mathbf{r}} \tau_{\mathbf{r}}$, where $n_{\mathbf{r}}$ is the number of visits to $\mathbf{r}$ up to time $t$. 
Although all the different $\tau_{\mathbf{r}}$ are i.i.d, the $\{ n_{\mathbf{r}} \}$ are correlated, like in our case of nearest-neighbor hopping on a lattice where $n_{\mathbf{r}}$ is very similar to the nearest-neighbour $n_{\mathbf{r'}}$.
By fixing the values of $\{ n_{\mathbf{r}} \}$ and  averaging over $\{\tau_{\mathbf{r}}\}$ (disorder averaging) the Laplace pair of $t$, i.e. $\langle e^{-u t}\rangle$, is $\sim e^{-A S_{\alpha} u^\alpha}$ where
\begin{equation}
    S_\alpha = \sum_{\mathbf{r}} (n_{\mathbf{r}})^\alpha.
    \label{eq: salpha definition}
\end{equation}
Since the Laplace pair of the one-sided L\'evy distribution $l_{\alpha,A,1}(\eta)$ is $\sim e^{-A u^{\alpha}}$ we obtain that $t\sim S_{\alpha}^{1/\alpha} \eta$, where $\eta$ is a random variable distributed according to $l_{\alpha,A,1}(\eta)$.
This connection between $t$, $\eta$ and fixed $S_\alpha$ allows to obtain the PDF of $S_\alpha$ for fixed $t$, $\mathcal{N}_t\left(\mathcal{S}_\alpha\right)$, by  changing variables from $\eta=t/{S_\alpha}^{1/\alpha}$ to $S_\alpha = (t/\eta)^\alpha$. Therefore,
$\mathcal{N}_t\left(\mathcal{S}_\alpha\right) \sim \frac{t}{\alpha}\left(\mathcal{S}_\alpha\right)^{-1 / \alpha-1} l_{\alpha, A, 1}\left[\frac{t}{\left(\mathcal{S}_\alpha\right)^{1 / \alpha}}\right]
$.
The explicit form of $\mathcal{N}_t\left(\mathcal{S}_\alpha\right)$ allows performing the first,  what is commonly called, subordination~\cite{klafter2011first} and express the disorder averaged $\langle P({\mathbf{r}},t)\rangle$ by using $S_\alpha$ as the local time of the process. 
Namely, for the conditional probabilty $P_{S_\alpha}({\mathbf{r}})$ of finding  the particle at position $\mathbf{r}$ for a given $S_\alpha$ (i.e., at ``time" $S_\alpha$), the law of total probability yields
\begin{equation}
    \label{eq: first subordination}
    \langle P(\mathbf{r}, t)\rangle=\sum_{S_\alpha} P_{S_\alpha}(\mathbf{r}) \mathcal{N}_t\left(S_\alpha\right).
\end{equation}
The second subordination is applied to $P_{S_\alpha}(\mathbf{r})$
We use the number of jumps, $N$, to represent $P_{S_\alpha}(\mathbf{r})$ in terms of $W_N(r)$, the probability to find the particle at ${\mathbf{r}}$ after $N$ jumps, and $\mathcal{G}_{S_\alpha, \mathbf{r}}(N)$ the probability of different values of $N$ for a prescribed $S_\alpha$ and $\mathbf{r}$. The law of total probability yields $P_{S_\alpha}(\mathbf{r})=\sum_{N=0}^{\infty} W_N(\mathbf{r}) \mathcal{G}_{S_\alpha, \mathbf{r}}(N)$ and then according to Eq.~\eqref{eq: first subordination}
\begin{equation}
\label{eq: second subordination}
    \langle P(\mathbf{r}, t)\rangle=\sum_{S_\alpha} \sum_{N=0}^{\infty} W_N(\mathbf{r}) \mathcal{G}_{S_\alpha}(N, \mathbf{r}) \mathcal{N}_t\left(S_\alpha\right).
\end{equation}
When $t\to \infty$, the probability of small $S_\alpha$ is negligible and for large $S_\alpha$, when the probability of eventual return to the origin $Q_0$ is $<1$, it was shown that~\cite{burov2017quenched}, $\mathcal{G}_{S_\alpha}(N, \mathbf{r}) \to \delta\left(S_\alpha-\Lambda N\right)$ where
\begin{equation} 
\label{eq: Lambda theory}
\Lambda=\left[\left(1-Q_0\right)^2 / Q_0 \right] L i_{-\alpha}\left(Q_0\right),
\end{equation}
$L i_{a}(b)=\sum_{j=1}^{\infty} b^{j} / j^{a}$ is the Polylogarithm function \cite{abramowitz1972handbook}.
$Q_0$
is computed when the spatial process is treated as a function of $N$.
Therefore, for QTM where the spatial process is defined solely by the jump probabilities $p(\mathbf{r}^{\prime} - \mathbf{r})$, Eq.~\eqref{eq: second subordination} yields
\begin{equation} 
\label{eq: PDF QTM}
    \langle P(\mathbf{r}, t)\rangle \sim \sum_N W_N(\mathbf{r}) \frac{t / \Lambda^{1 / \alpha}}{\alpha N^{\frac{1}{\alpha} - 1}}  l_{\alpha, A, 1}\left(\frac{t / \Lambda^{1 / \alpha}}{N^{1 / \alpha}}\right) .
\end{equation}
Equation~\eqref{eq: PDF QTM}, first obtained in~\cite{burov2017quenched}, presents the disorder averaged propagator of QTM in terms of the spatial process on a lattice as a function of $N$, and a transformation from $N$ to $t$.
Two points are in place: (I) The distribution $W_N(r)$ is defined by the jump probabilities $p(\mathbf{r}'-\mathbf{r})$ and found by the standard techniques for a random walk (RW) on a lattice~\cite{weiss1994aspects}. (II) For $\Lambda = 1$ Eq.~\eqref{eq: PDF QTM} displays the propagator for the annealed version of the disorder 
(CTRW)~\cite{bouchaud1990anomalous}. Therefore, $\Lambda$, quantifies the difference between quenched and annealed disorder as a function of $Q_0$ (Eq.~\eqref{eq: Lambda theory}) and depends on the geometry and the external force.
Below, we utilize Eq.~\eqref{eq: PDF QTM} and compute the response to external constant force $F$ acting on a single particle in a $2$-dimensional channel of width $w$. For this purpose, we first compute the average position along the longitudinal axis of the channel $\hat{x}$.

\begin{figure}[t]
\centering
\includegraphics[width=0.45\textwidth]{"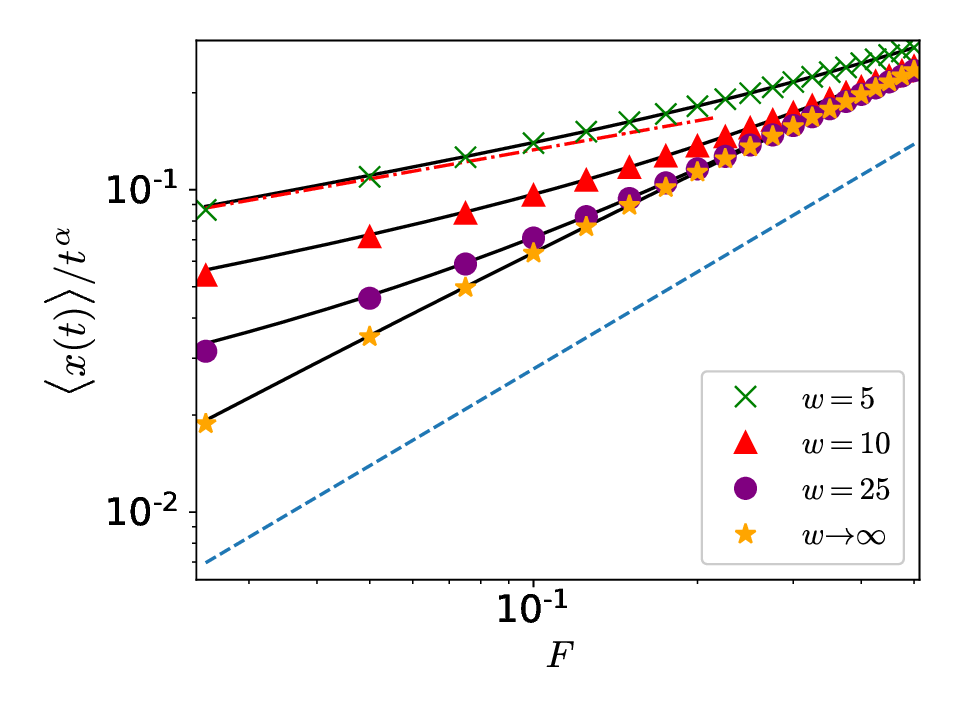"}
\caption{The average displacement in the direction of application of external force $F$. $\langle x(t) \rangle$ is growing as channel width $w$ is reduced.
Solid lines display theoretical prediction (Eq.~\eqref{eq: average position}). The red dash-dot line is the approximation for small-$F$ (Eq.~(\ref{eq: small F response})).  Symbols are simulation results averaged over $3\times10^6$ trajectories, $\alpha=0.3$, $A=1$ and $t=10^{14}$ (a.u).
The blue dashed line is obtained for the equivalent annealed disorder (CTRW) system.
}
\label{Fig:1}
\end{figure}

\textit{The average displacement.}
The motion occurs on top of a lattice in a two-dimensional channel and is unrestricted in the $\hat{x}$ direction.
The width of the channel, in the $\hat{y}$ direction, is $w$, which is also the number of sites across  $\hat{y}$ (the lattice spacing is $a=1$). Due to the translational invariance of the spatial process and transition probabilities $p(\mathbf{r}-\mathbf{r'})$, we use periodic boundary conditions for $\hat{y}$. 
The case of reflecting boundary conditions will be addressed below. 
The strength of the force $f$, applied only along $\hat{x}$, is characterized by the dimensionless parameter $F= a f/ T$, where $k_B$ is set to $1$. 
The transition probabilities $p(\mathbf{r}-\mathbf{r'})$ allow transitions only to the nearest neighbors on the square lattice. Namely, $p_\rightarrow$ ($p_\leftarrow$) is the probability for a single jump to the right (left) along $\hat{x}$ and $p_\uparrow$ ($p_\downarrow$) is the probability for a single jump up (down) along $\hat{y}$.  The detailed balance condition dictates that $p_\rightarrow / p_{\leftarrow}=e^F$ and $p\uparrow / p_\downarrow = 1$. 
Therefore, due to the normalization condition $p_\uparrow+p_\downarrow+p_\leftarrow+p_\rightarrow=1$, we obtain that $p_\rightarrow =B e^{F/2}$, $p_\leftarrow = B e^{-F/2}$ and $p_\uparrow=p_\downarrow=B=1 /[2 \cosh (F / 2)+2]$. 
We are interested in the mean position $\langle x(t) \rangle
=\sum_{\mathbf{r}} x \langle P(\mathbf{r},t)\rangle$.
After one single jump the average displacement along $\hat{x}$ is $p_\rightarrow-p_\leftarrow=\tanh(F/4)$, therefore after $N$ jumps the average displacement is $\sum_{\mathbf{r}}x W_N(\mathbf{r})=N \tanh(F/4) $. 
Then according to Eq.~\eqref{eq: PDF QTM} $\langle x(t)\rangle = \sum_N N\tanh(F/4) \frac{t / \Lambda^{1 / \alpha}}{\alpha N^{\frac{1}{\alpha} - 1}}  l_{\alpha, A, 1}\left(\frac{t / \Lambda^{1 / \alpha}}{N^{1 / \alpha}}\right)$. 
We take the limit $t\to\infty$, replace the summation by integration~\cite{bouchaud1990anomalous} and use the relation $\int_0^{\infty} y^q l_{\alpha, A,1}(y) d y=A^{q/\alpha}\Gamma(1-q / \alpha) / \Gamma(1-q)$ for $q / \alpha<1$~\cite{barkai2001fractional} and obtain  the average displacement in $\hat{x}$
\begin{equation} \label{eq: average position}
        \langle x(t)\rangle \sim \frac{\tanh (F/4)}{A \Gamma[1+\alpha]} 
        \frac{Q_0}{(1-Q_0)^2 \mathrm{Li}_{-\alpha}(Q_0)} t^\alpha.
\end{equation}
Eq.~\eqref{eq: average position} shows that the average displacement is anomalous in time $\sim t^\alpha$.
Such departure from the Einstein relation that predicts a linear dependence on time is a direct consequence of diverging mean waiting times, and for the annealed disorder was termed as Generalized Einstein relation~\cite{Barkai1998}.
The return probability $Q_0$ (Eq.~\eqref{eq: Lambda theory}) depends on geometry, jump probabilities, and $F$.
To finalize the calculation of $\langle x(t)\rangle$ we find the explicit form of $Q_0$. 

\begin{figure*}
\includegraphics[width=0.95\textwidth]{"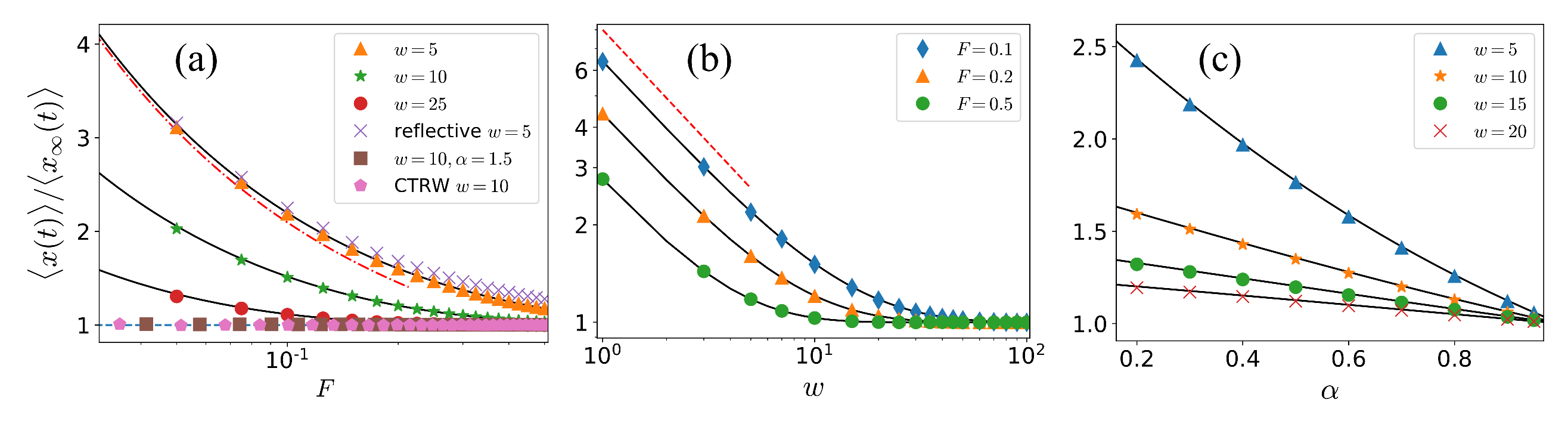"}
\caption{\label{fig:combined}
Enhancement of mobility in a channel of width $w$ ($\langle x(t)\rangle$) with respect to mobility in unrestricted $2d$ geometry ($\langle x_\infty(t)\rangle$), as a function of external force $F$ (panel {\bf(a)}), width $w$ ({\bf(b)}) and $\alpha=T/T_g$ ({\bf (c)}).
In all panels, Eq.~\eqref{eq: average position} (combined with Eq.~\eqref{eq: Q bounded system}) is displayed by solid black lines, and the symbols are the results of numerical simulations.
{\bf(a)}: The red dash-dot line is the small-$F$ expansion while the leading term is provided by Eq.~(\ref{eq: speed-up}) (see SM for full expression).
The $\times$ presents simulation results for reflective boundary conditions and $w=5$ that almost follows the corresponding case with periodic boundary conditions $\triangle$.
The blue dashed line represents the case when no enhancement was detected: for annealed disorder (CTRW) and QTM with a finite average dwell times ($\alpha > 1$). For all cases $t=10^{14}$ (a.u.) and except $\square$, $\alpha=0.3$. 
{\bf(b)}: The red dashed line indicates the $w^{\alpha-1}$ scaling, $\alpha = 0.3$, and $t=10^{14}$ (a.u.).
{\bf(c)}: $F=0.1$ and $t^\alpha=4.2$ (a.u.). For all panels, $A=1$.
}
\end{figure*}

\textit{The  return probability $Q_0$} is computed in terms of $f_N(\mathbf{0})$, the first return probability to the origin after $N$ steps. 
Namely, $Q_0=\sum_{N=0}^\infty f_N(0)$. 
The probability $f_N(\mathbf{0})$ determines the probability $W_N(\mathbf{0})$ since according to the renewal equation~\cite{weiss1994aspects}, $W_N(\mathbf{0})=\delta_{N,0}+\sum_{i=1}^N f_N(\mathbf{0})W_{N-i}(\mathbf{0})$, which yields for the generating function of the first return probability, $\tilde{f}_z(\mathbf{0})=\sum_{N=0}^\infty f_N(\mathbf{0})z^N$, the result $\tilde{f}_z(\mathbf{0})=1-1/\tilde{W}_z(\mathbf{0})$, where $\tilde{W}_z(\mathbf{0})=\sum_{N=0}^\infty W_N(\mathbf{0})z^N$. By noting that $Q_0=\tilde{f}_1(\mathbf{0})$ the connection between $Q_0$ and $W_N(\mathbf{0})$ is finally established~\cite{weiss1994aspects}, namely $Q_0=1-1/\tilde{W}_{z=1}(\mathbf{0})$. 
Since  $W_N(\mathbf{r})$ is a convolution of $N$ random variables, i.e., steps, its Fourier transform is the $N$th power of the single-step characteristic function 
$\lambda(\mathbf{k})=e^{ik_x}p_\rightarrow+e^{-ik_x}p\leftarrow+e^{ik_y}p_\uparrow+e^{-ik_y}p_\downarrow$ 
and therefore  $\tilde{W}_{z=1}(\mathbf{0})=\frac{1}{4\pi^2}\int_{-\pi}^{\pi}\int_{-\pi}^{\pi}\frac{1}{1-\lambda(\mathbf{k})}dk_x dk_y$. 
In the supplemental material (SM) we show how this integral is computed and eventually obtain the explicit form of the return probability $Q_0$
\begin{equation} 
\label{eq: Q bounded system}
    Q_0 = 1 - \frac{w\Big/[1+\cosh(\frac{F}{2})]}{\displaystyle\sum_{m=0}^{w-1}1\Big/\sqrt{\left[1+\cosh\left(\frac{F}{2}\right)-\cos \left(\frac{2 \pi m}{w}\right)\right]^2-1}}.
\end{equation}
When $w\to\infty$, the result in Eq.~\eqref{eq: Q bounded system} converges to the known result~\cite{montroll1979enriched} for an unbounded $2$-dimensional square lattice $\lim_{w\to\infty}Q_0 = 1-1\Big/\left[\frac{2}{\pi}\mathbf{K}\left(\frac{4}{(1+\cosh(\frac{F}{2}))^2}\right)\right]$, where $\boldsymbol{K}(k) = \int_0^{\pi /2} d\gamma \left( 1 - k \sin ^2 \gamma \right)^{-1/2} $ is the complete elliptic integral of the first kind~\cite{abramowitz1972handbook}.

Equations~(\ref{eq: average position} - \ref{eq: Q bounded system}) 
provide $\langle x(t) \rangle$ 
as a function of time,  arbitrarily large external force $F$ and the width of the channel $w$.
For small forces Eq.~\eqref{eq: average position} yields (see SM)
\begin{equation} \label{eq: small F response}
    \langle x(t) \rangle \sim \frac{w^{\alpha-1}}{A \Gamma^2[1+\alpha]}  \left(\frac{F}{4}\right)^\alpha t^\alpha,
\end{equation}
while $4/F \gg w / \sqrt{2}$ and $w$ is an integer $ \geq 1$. 
Our main result is immediately apparent from 
Eq.~\eqref{eq: small F response}: $\langle x(t) \rangle$ unexpectedly decays with growing channel width $w$! The dependence is $w^{\alpha-1}$, meaning the motion is faster for narrow channels than for wide channels. In Fig.~\ref{Fig:1} an excellent agreement between this analytical result and numerical simulation is displayed. In addition, we observe that the dependence on $F$ is complex and non-linear as indicated through Eq.\eqref{eq: Q bounded system} and seen in Fig.~\ref{Fig:1}. For small $F$ the dependence is simplified to $\sim F^\alpha$.
Namely, the strong disorder and it's quenched nature that impose prolonged correlations make the usual assumption of linear response inapplicable in the quasi $2$d, as was found previously for $1$d QTM ~\cite{bouchaud1990anomalous,monthus2004nonlinear,burov2017quenched,shafir2022case}.
We note that for the case of strong annealed disorder, (CTRW), the dependence on $F$ (when $F\to 0$) is linear~\cite{shafir2022case} (see the dashed line in Fig.~\ref{Fig:1}).

To emphasize the mobility enhancement due to the channel width constraint, we calculate (see SM) the ratio of the $\langle x(t) \rangle$ for a given $w$, and the average displacement for unrestricted $2$d motion, $\langle x_{\infty}(t) \rangle$, i.e., $w\to\infty$.
For $F\to 0$ we obtain
\begin{equation} \label{eq: speed-up}
  \langle x(t)\rangle\Big/\langle x_{\infty}(t)\rangle
  \sim
  \left[(w/4\pi) F \ln \left(128 / F^2\right)\right]^{\alpha-1},
\end{equation}
implying that imposing a geometrical constraint enhances the transport, and the stronger the constraint (narrower channel), the larger the enhancement! 
Note that the logarithmic term in $F$ enters Eq.~\eqref{eq: speed-up} due to the critical properties of $Q_0$ in unrestricted $2$d (see~\cite{shafir2022case} and SM for details). 
In Fig.\ref{fig:combined} we present the excellent agreement between the analytical results for $\langle x(t)\rangle\Big/\langle x_{\infty}(t)\rangle$ and numerical simulations when explored as a function of $F$, $w$ and $\alpha$. 
The enhancement associated with geometrical restriction repeats itself in all the presented cases and is preserved for smaller times (see SM). 
Fig.~\ref{fig:combined}(a) shows that the effect disappears  ($\langle x(t)\rangle\Big/\langle x_{\infty}(t)\rangle=1$)  when the disorder is not strong ($\alpha>1$), or if the disorder is not quenched. 

In our derivation, we assumed periodic boundary conditions in the channel. In Fig.~\ref{fig:combined}(a), we show (numerically) that for reflecting boundary conditions, the obtained effects of non-linear dependence on $F$ and transport enhancement due to geometrical restriction are preserved. Additional details are provided in SM, and we intend to address this issue in future work. 
Equation~\ref{eq: small F response} summarizes the unconventional effect of strong and quenched disorder. The regular expectation for $\langle x(t)\rangle$ is $\langle x(t)\rangle=\mu F t$ where $\mu$ is the mobility. Strong disorder modifies the temporal dependence and the regular Einstein relation. Quenchness breaks linear response and introduces the non-linear dependence on $F$, and here we have shown that the properties of the mobility $\mu$ are counterintuitive. First of all, the mobility $\mu$ for QTM is anomalous since it can't be defined as $\langle x(t) \rangle/tF$, but rather as $\mu=\langle x(t) \rangle/t^\alpha F^\alpha$. From Eq.~\eqref{eq: small F response} $\mu=w^{\alpha-1}\big/4^\alpha A \Gamma^2[\alpha+1]$ while $\alpha=T/Tg<1$. The mobility is enhanced as the channel width $w$ decreases. While the expectation is that additional pathways, which start to appear with relaxed geometrical constraints, will lead to a speed-up of the transport~\cite{Schwarcz2022},  we observe an opposite behavior. In the presence of strong and quenched disorder, stricter geometrical constraints improve mobility.

Mathematical reasons for such counterintuitive enhancement are rooted in the properties of the local time $S_\alpha$, transformation constraint $\Lambda$, and geometrical dependence of $Q_0$. 
The intuition behind the found effect is based on what is known as the ``big jump principle"~\cite{Vezani2019,Vezanni2020,Singh2023}. When scale-free  waiting times (Eq.~\eqref{eq: waiting time PDF}) are in play, it is not the accumulation of many events but rather a single maximal event that governs the overall behavior. Naturally, when such a single event is excluded, for example, by replacing the site with maximal waiting time by significantly shorter $\tau$, it will lead to faster transport~\cite{holl2023controls}. Our results suggest that narrowing the channel width decreases the number of possible sites the particle will sample during transport and effectively modifies this single dominant arrest time. The quenched nature of the disorder is a crucial ingredient for this to work. 
A further in-depth analysis and experimental research of this phenomenon is warranted. 
We expect that such enhancement will be useful to optimize transport in a channel media relevant for applications in nanotechnology and nanomedicine~\cite{malgaretti2019special} and transport in porous media~\cite{holl2023controls}.

\begin{acknowledgments} This work was supported by the  Israel Science Foundation Grant No. 2796/20.
\end{acknowledgments}

\section{Supplemental Material}

Supplemental material includes 
(\ref{SM: one}) calculation of the return probability with periodic boundary conditions. 
(\ref{SM: two}) Analytical derivation of the response in the $F \to 0$ limit. 
(\ref{SM: three}) Analytical derivation of the mobility enhancement: the ratio of the $\langle x(t) \rangle$ for a given $w$, and the average displacement for unrestricted $2$d motion, $\langle x_{\infty}(t) \rangle$, i.e., $w\to\infty$.
(\ref{SM: four}) calculation of the return probability with reflective boundary conditions. 
(\ref{SM: five}) Numerical investigation of the mobility enhancement effect as a function of time.

\setcounter{equation}{0}
\renewcommand{\theequation}{S.\arabic{equation}}

\subsection{The return probability with periodic boundary conditions \label{SM: one}}
This section provides the full analytical derivation of the return probability of a walker on a two-dimensional lattice. The geometry is a channel (integer width $w \geq 1$), and the motion obeys periodic boundary conditions at channel walls $y=0$ and $y=w$.

The return probability $Q_0$ is found by the means of generating functions~\cite{weiss1994aspects}.
We use the probability of first return to the starting point ($\mathbf{r}=0$) after $N$ steps, i.e., $f_N (0)$, and represent $Q_0$ as  
$Q_0=\lim_{z\rightarrow1} \left( \sum_{N=0}^{\infty} f_N (0) z^N \right)$.
$W_N (0)$ is the probability to find the RW at position $\mathbf{r}=0$ after $N$ steps. 
The quantities $f_N (0)$ and $W_N(0)$ are related due to the renewal equation~\cite{weiss1994aspects}, $W_N(\mathbf{0})=\delta_{N,0}+\sum_{i=1}^N f_N(\mathbf{0})W_{N-i}(\mathbf{0})$.
Multiplying this relation by $z^N$ and summing on all possible $N$'s we receive the connection
\begin{equation}
    \sum_{N=0}^{\infty} f_N (0) z^N = 1-\frac{1}{\sum_{N=0}^{\infty} W_N (0) z^N},
\end{equation}
which means that
\begin{equation} \label{eq: Q_0 definition}
    Q_0= 1-\lim_{z\rightarrow1}\frac{1}{\left(\sum_{N=0}^{\infty} W_N (0) z^N\right)}.
\end{equation}
Where in the denominator we have the generating function of the unbounded system $\tilde{W}_z(\mathbf{r})=\sum_{N=0}^{\infty} W_N(\mathbf{r}) z^N$ , the $z$-transform of the positional probability of the walker $W_N(\mathbf{r})$ after $N$ steps.
The generating function is evaluated for a random walk commencing at $\mathbf{r} = 0$. 
The summation is often solved by making use of $\lambda(\mathbf{k})$, the characteristic function ($\mathbf{k}$-space transform) of a single jump $\Delta \mathbf{r}$ distribution $p(\Delta \mathbf{r})$ defined by $\lambda(\mathbf{k}) = \langle e^{i \mathbf{k} \cdot \Delta \mathbf{r}} \rangle = \sum_{\Delta \mathbf{r}} p(\Delta \mathbf{r})e^{i \mathbf{k} \cdot \Delta \mathbf{r}} $. Then, by using the convolution theorem we have $W_N(\mathbf{k})=\lambda(\mathbf{k})^N$ and $\tilde{W}_z(\mathbf{r})$ can now be found by transforming back to $\mathbf{r}$-space:
\begin{eqnarray} \label{eq: generating function integral form}
\tilde{W}_z(\mathbf{r}) &=& \frac{1}{(2 \pi)^2} \int_{-\pi}^{\pi} \int_{-\pi}^{\pi } \left(\sum_{N=0}^{\infty} \lambda(\mathbf{k})^N z^N \right) \mathrm{d}^2 \mathbf{k} \nonumber \\
&=& \frac{1}{(2 \pi)^2} \int_{-\pi}^{\pi} \int_{-\pi}^{\pi } \frac{ e^{-i \mathbf{k} \cdot \mathbf{r}}}{1-z \lambda(\mathbf{k})}  \mathrm{d}^2 \mathbf{k}.
\end{eqnarray}
Here $\lambda(\mathbf{k})$ is
\begin{eqnarray} \label{eq: characteristic function}
\lambda(\mathbf{k})=&&2 B \big[\cos \left(k_x\right) \cosh (F / 2)\nonumber\\
&&+i \sin \left(k_x\right) \sinh (F / 2)+\cos \left(k_y\right)\big],
\end{eqnarray}
and $B = 1/ [2 \cosh (F/2) + 2]$ is the normalization of the transition probabilities of a single step.

To evaluate the return probability $Q_0^*$ for the bounded system in quasi $2$d (of a channel geometry) with a finite lattice $\Omega$ we use the solution of the infinite lattice in $2$d and invoke periodic boundary conditions at the channel walls~\cite{hughes1995random}:
\begin{equation}
    W^*_N(\mathbf{r}) = \sum_m W_N (\mathbf{r} + (0, w m)),
\end{equation}
where $w$ is the width of the channel and $m$ is an integer.
This implies for the generating function of the bounded system 
\begin{eqnarray} \label{eq: bounded generating function periodic}
    \tilde{W}_z^*(\mathbf{r}) &=& \sum _m \tilde{W}_z(\mathbf{r} + m(0,w)) \\ \nonumber
    &=& \sum_m \frac{1}{(2 \pi)^2} \int_{-\pi}^{\pi} \int_{-\pi}^{\pi } \frac{e^{-i \mathbf{k} \cdot (x, y+m w)}}{1-z \lambda(\mathbf{k})} \mathrm{d}^2 \mathbf{k}.
\end{eqnarray}
We make use of the well-known representation of the delta function
\begin{equation} \label{eq: delta function representation}
    \sum_{m=-\infty}^{\infty} e^{-i m k} = 2 \pi \sum_{m=-\infty}^{\infty} \delta(k - 2 \pi m),
\end{equation}
hence
\begin{eqnarray}
    \sum_m e^{-i m w k_y} &=& 2 \pi \sum_m \delta\left(w k_y-2 \pi m\right) \\ \nonumber
    &=& \frac{2 \pi}{w} \sum_m \delta\left(k_y-\frac{2 \pi m}{w}\right).
\end{eqnarray}
The integrand in Eq.~(\ref{eq: bounded generating function periodic}) is periodic, so that we may replace the integration region $B=[-\pi , \pi]^2$ by the region $[-\epsilon, 2 \pi-\epsilon]^2$. We choose $\epsilon$ such that $0<\epsilon<\frac{2 \pi}{w}$, and interchange order of summation and integration. The singularity of the delta function $\delta (k_y - \frac{2 \pi m}{w})$ is in the integration region if and only if $(0,m) \in \Omega$, therefore
\begin{equation}
    \begin{aligned}
        \tilde{W}_z^*(\mathbf{r}) &=& \frac{1}{w} \frac{1}{2 \pi} \int_{-\pi}^\pi d k_x \sum_{m=0}^{w-1} \frac{e^{-i \mathbf{r} \cdot\left(k_x, \frac{2 \pi m}{w}\right)}}{1-z \lambda \left(\left(k_x, \frac{2 \pi m}{w}\right)\right)}.
        \end{aligned}
\end{equation} 
To obtain the return portability, we need to evaluate this function at $\mathbf{r}=0$ 
\begin{eqnarray} \label{eq: SM pr}
    \tilde{W}_z^*(0)=\frac{1}{2 \pi w} \sum_m \int_{-\pi}^\pi \frac{d k_x}{1-z \lambda\left(\left(k_x, \frac{2 \pi m}{w}\right)\right)}.
\end{eqnarray}
We first solve the integral in the sum in Eq.\eqref{eq: SM pr} for arbitrary $\mathbf{k}=(k_x, k_y)$, i.e. with $\lambda(k_x, k_y)$. 
For that purpose, 
we use the relation 
\begin{eqnarray}
    \int_0^\infty e^{-s a} ds = 1 / a \quad (a>0),
\end{eqnarray}
and find
\begin{eqnarray}
    U&=&\int_{-\pi}^\pi \frac{d k_x}{1-z \lambda(\mathbf{k})} =\int_{-\pi}^\pi d k_x \int_0^{\infty} d s e^{-s(1-z \lambda(\mathbf{k}))} \nonumber \\
    &=& \int_0^\infty e^{-s} ds \int_{-\pi}^{\pi} dk_x e^{s z \lambda(\mathbf{k})}.
\end{eqnarray}
Now we substitute the value of $\lambda(\mathbf{k)}$ from Eq.~(\ref{eq: characteristic function}),
\begin{eqnarray}
    && U=e^{2 s z B \cos \left(k_y\right)} \int_0^{\infty} e^{-s} d s \int_{-\pi}^\pi d k_x  \\ \nonumber
    && \times \exp \left[ 2 s z B\Big(\cos (k_x) \cosh (F / 2)+i \sin \left(k_x\right) \sinh (F / 2 )\Big) \right]
\end{eqnarray}
and make use of the relation (Eq. (64) from \cite{montroll1973random})
\begin{eqnarray}
    &&\int_{-\pi}^\pi \frac{\mathrm{d} k}{2 \pi} \exp [-i k m] \exp [C_1 \cos (k)+ i C_2 \sin (k)]= \nonumber \\
    &&\left[\frac{C_1+C_2}{\sqrt{C_1^2-C_2^2}}\right]^m I_m\left(\sqrt{C_1^2-C_2^2}\right),
\end{eqnarray}
where $I_m(\dots)$ is a modified Bessel function of the first kind of integer order $m$~\cite{abramowitz1972handbook}. 
With $m=0, C_1=2 s z B \cosh (F/2)$ and $C_2 = 2s z B \sinh (F/2)$, we obtain
\begin{eqnarray}
    U=\int_0^{\infty} d s e^{-s} e^{2 s z B \cos \left(k_y\right)}2 \pi I_0\left(2 s z B\right).
\end{eqnarray}
Substituting this back into the generating function yields
\begin{eqnarray}
    \tilde{W}_z^*&(&0)= \\ \nonumber
    &&\frac{1}{w} \sum_{m=0}^{w-1} \int_0^{\infty} d s e^{-s} I_0\left(2 s z B\right) \exp \left[2 s z B \cos \left(\frac{2 \pi m}{w}\right)\right].
\end{eqnarray}
To evaluate the integral we use the formula (Eq.~(4), page 695 in \cite{gradshteyn2014table}),
\begin{eqnarray}
    \int_0^\infty e^{-s b} I_0 (s a) = \frac{1}{\sqrt{b^2 - a^2}}, \quad a<b
\end{eqnarray}
with $a=2 z B$ and $b=1 - a \cos \left( \frac{2 \pi m}{w} \right)$, and obtain
\begin{eqnarray}
    \tilde{W}_z^*&(&0)= \\ \nonumber
    &&\frac{1}{w} \sum_{m=0}^{w-1}\left[\left(1-2 B z \cos \left(\frac{2 \pi m}{w}\right)\right)^2-4 z^2 B^2\right]^{-1/2}.
\end{eqnarray}
Finally,  the result for the return probability in a channel with periodic boundary conditions is 
\begin{eqnarray} \label{eq: return probability SM}
    Q_0^* &=& 1 - 1/\tilde{W}_1^*(0) \\ \nonumber
    &=& 1 - \frac{w}{\sum_{m=0}^{w-1}\left[\left(1-2 B \cos \left(\frac{2 \pi m}{w}\right)\right)^2-4 B^2\right]^{-\frac{1}{2}}}.
\end{eqnarray}

\subsection{The response for small forces \label{SM: two}}
In this section we derive the response along parallel direction to the walls for small force $F$.
We first expand the return probability $Q_0$ of a RW in a channel for small forces, then plug it into equation (7) from the main text:
\begin{equation} \label{eq: first moment SM}
    \langle x(t)\rangle \sim \frac{\tanh(F/4)}{A \Gamma [1+\alpha]} \frac{Q_0}{(1-Q_0)^2 \text{Li}_{-\alpha}(Q_0)} t^\alpha.
\end{equation}
The denominator in the return probability $Q_0$ in Eq.~(\ref{eq: return probability SM}) is expanded for small $F$,
\begin{eqnarray} \label{eq: Q0 denomiator}
    && \sum_{m=0}^{w-1}\left[\left(1-2 B \cos \left(\frac{2 \pi m}{w}\right)\right)^2-4 B^2\right]^{-1 / 2} \\ \nonumber
    \approx && \sum_{m=0}^{w-1}\left[\frac{1}{4}(\xi-2)^2-\frac{1}{4}-\left((\xi-1)^2-2\right) \frac{F^2}{32}\right]^{-1 / 2},
\end{eqnarray}
where we set for convenience $\xi = \cos \left(\frac{2 \pi m}{w}\right)$.
We rewrite Eq.~\eqref{eq: Q0 denomiator} as
\begin{eqnarray} \label{eq: Q0 denomiator 2}
    \frac{4}{F}+\left(1-\delta_{w, 1}\right) \left(\frac{1}{4}(\xi-2)^2-\frac{1}{4}\right)^{-1 / 2} \\ \nonumber
    \times \sum_{m=1}^{w-1}\left[1-\frac{(a-1)^2-2}{\frac{1}{4}(\xi-2)^2-\frac{1}{4}} \frac{F^2}{32}\right]^{-1/2},
\end{eqnarray}
where $\delta_{i,j}$ is Kronecker's delta function.
Expanding again for small $F$ the argument in the sum and rearranging as a series in powers of $F$, Eq.~\eqref{eq: Q0 denomiator 2} (the denominator of $Q_0$ in Eq.~(\ref{eq: return probability SM})) becomes
\begin{eqnarray}
    \frac{4}{F}+\left(1-\delta_{w, 1}\right) \sum_{m=1}^{w-1}\left(\frac{1}{4}(\xi-2)^2-\frac{1}{4}\right)^{-1 / 2} \\ \nonumber
    +\left(1-\delta_{w, 1}\right) \sum_{m=1}^{w-1} \frac{(\xi-1)^2-2}{\left(\frac{1}{4}(\xi-2)^2-\frac{1}{4}\right)^{3 / 2}} \frac{F^2}{32} + \mathcal{O}(F^4).
\end{eqnarray}
For small $F$ the first two terms are the dominant ones. We check when does the second term (the force free) can also be neglected. The maximum value in the sum is achieved when $2 \pi m / w = \pi$, therefore the sum has an upper bound,
\begin{eqnarray}
    \left\| \sum_{m=1}^{w-1}\left[\frac{1}{4}\left(\cos \left(\frac{2 \pi m}{w}\right)-2\right)^2-\frac{1}{4}\right]^{-1 / 2} \right\| \\ \nonumber
    \leq \sum_{m=1}^{w-1}\left[\frac{1}{4}(\cos (\pi)-2)^2-\frac{1}{4}\right]^{-1 / 2} = \frac{w}{\sqrt{2}}.
\end{eqnarray}
So as long as $4/F \gg w/\sqrt{2}$, we are safe to assume that the value found for $Q_0$ in Eq.~(\ref{eq: return probability SM}) for the channel system obeys
\begin{equation}
    \sum_{m=0}^{w-1}\left[\left(1-2 \xi \cos \left(\frac{2 \pi m}{w}\right)\right)^2-4 \xi^2\right]^{-1 / 2} \approx \frac{4}{F}.
\end{equation}
The return probability for small $F$ and $4/F \gg w/\sqrt{2}$ now becomes
\begin{equation}
    Q_0 = 1 - \frac{wF}{4}.
\end{equation}
We use this result in Eq.~(\ref{eq: first moment SM}) together with the asymptotic relation $L i_{-\alpha}(1-\epsilon) \sim \Gamma[1+\alpha] \epsilon^{-\alpha-1}$ in the limit of $\epsilon \to 0$ \cite{abramowitz1972handbook} and obtain
\begin{eqnarray}
    \langle x(t) \rangle &\approx& \frac{(F / 4)}{A\Gamma^2[1+\alpha]}\left[\left(\frac{w F}{4}\right)^{\alpha-1}-\left(\frac{w F}{4}\right)^\alpha\right] t^\alpha \nonumber \\
    &\approx& \frac{1}{A \Gamma^2[1+\alpha]} \left(\frac{1}{w} \right)^{1-\alpha} \left(\frac{F}{4}\right)^\alpha t^\alpha.
\end{eqnarray}

\subsection{Analytical derivation for the transport enhancement vs. the unbounded system \label{SM: three}}
In this section we show the analytical derivation of the mobility enhancement: the ratio of the $\langle x(t) \rangle$ for a given $w$, and the average displacement for unrestricted $2$d motion, $\langle x_{\infty}(t) \rangle$, i.e., $w\to\infty$.
The average position $\langle x_\infty (t)\rangle$ in the unbounded system will be provided again by expanding the return probability $Q_0$ for small forces.
In Ref.~\cite{montroll1979enriched}) (see Eqs.~(6.43-6.44), the return probability $Q_0$ for the unbounded system was found to be
\begin{eqnarray}
Q_0 &=& 1 - 1 / \left[ \int_0^{\infty} ds e^{-s} \left[ I_0 (2 s B) \right]^2 \right] \\ \nonumber
&=& 1 - 1 / \left[ \frac{2}{\pi} \boldsymbol{K}\left(16 B^2 \right) \right],
\end{eqnarray}
where $\boldsymbol{K}(k) = \int_0^{\pi /2} d\gamma \left( 1 - k \sin ^2 \gamma \right)^{-1/2} $ is the complete elliptic integral of the first kind. 
Using the approximation $(2 / \pi) \mathrm{K}(z^2) \approx (1/ \pi) \text{Log} (8 / (1-z))$ for the complete elliptic integral of the first kind as $z \to 1$ from Ref.~\cite{brummelhuis1988single} we obtain,
\begin{eqnarray}
    Q_0&=&1-1 /\left[\frac{2}{\pi} \boldsymbol{K}\left(16 \left(\frac{1}{2+2 \cosh(F/2)}\right)^2\right)\right] \nonumber \\
    &=&1- \pi / \left[ \ln (128 / F^2) \right].
\end{eqnarray}
In this case, the return probability decays much faster than the channel system ($Q_0 \approx 0.9$ already at $F\approx10^{-8}$), and we have to take more terms in the series expansion of $Li_{-\alpha}(1-\epsilon)$ around $\epsilon \to 0$. By using Mathematica we find,
\begin{eqnarray} \label{eq: Li approximation}
    Li_{-\alpha}(Q_0) && \sim \Gamma [\alpha +1] \left(\frac{\pi}{\log \left(\frac{128}{F^2}\right)}\right)^{-\alpha } \\ \nonumber
    && \times \left(\frac{ \log \left(\frac{128}{F^2}\right)}{\pi }-\frac{1}{2} (\alpha +1)\right) +\zeta (-\alpha ),
\end{eqnarray}
where $\zeta(s)=\sum_{k=1}^{\infty} k^{-s}$ is the Riemann zeta function defined for $Re(s) > 1$ (for $Re(s)<1$ we can use Riemann's functional equation $\zeta(s)=2^s \pi^{s-1} \sin \left(\frac{\pi s}{2}\right) \Gamma[1-s] \zeta(1-s)$).
The approximation in Eq.~(\ref{eq: Li approximation}) has an excellent agreement of one percent error for values as large as $F \approx 0.1$.
Substitution of this expression into Eq.~(\ref{eq: first moment SM}) yields
\begin{eqnarray}
    \langle x_\infty(t) \rangle &&\sim \frac{F/4}{A \Gamma [1+\alpha]} \\ \nonumber
    &&\times \frac{\left(\xi\right)^{-2}-\left(\xi\right)^{-1}}{\Gamma[1+\alpha] \left(\left(\xi\right)^{-\alpha-1}-\frac{1}{2} (\alpha+1) \left(\xi\right)^{-\alpha} \right) + \zeta(-\alpha)} t^\alpha,
\end{eqnarray}
where $\xi = \frac{\pi}{\log \left(\frac{128}{F^2}\right)}$ and the subscript of $\infty$ signifies the solution for the boundary-free system.
The transport enhancement is now provided by the fraction:
\begin{eqnarray} \label{eq: mobility enhancement exact}
    \frac{\langle x(t) \rangle}{\langle x_\infty(t) \rangle} &&\sim \left(\frac{F}{4}\right)^{\alpha -1} \left(\frac{1}{w}\right)^{1-\alpha } \\ \nonumber
    &&\times \frac{\left(\frac{\log \left(\frac{128}{F^2}\right)}{\pi }-\frac{\alpha +1}{2}\right) \left(\frac{\log \left(\frac{128}{F^2}\right)}{\pi }\right)^{\alpha }+\frac{\zeta (-\alpha )}{\Gamma (\alpha +1)}}{\frac{\log ^2\left(\frac{128}{F^2}\right)}{\pi ^2}-\frac{\log \left(\frac{128}{F^2}\right)}{\pi }}.
\end{eqnarray}
The leading term in $F$ is 
\begin{equation}
\frac{\langle x(t) \rangle}{\langle x_\infty(t) \rangle} \sim \left(\frac{4 \pi}{w F \ln \left( \frac{128}{F^2} \right)}\right)^{1-\alpha}.
\end{equation}

\subsection{The return probability with reflective boundary conditions \label{SM: four}}
In this section, we evaluate the return probability with reflective boundary conditions using the method of images (See chapters 21.5.3 and 21.5.4 in \cite{riley1999mathematical}). We show that the answer is the same as with periodic reflective conditions (for trajectories starting at $\mathbf{r} = 0$).

For reflective channel walls, i.e Neumann boundary conditions, the derivative of the generating function perpendicular to the channel walls has to be zero to ensure no flux flow. This condition can be imposed by looking at the solution $\tilde{W}_z^* (\mathbf{r})$ as a linear combination of the general solution $\tilde{W}_z (\mathbf{r})$. To ensure the derivative is zero at the channel walls, for each point $y_0$ in the interior ($0<y_0<w$), we place an image charge reflected by each of the channel walls with the same sign, resulting in two infinite sets of image charges located at $\{-y_0 + (2m+1)w\}$ and $\{y_0+2 m w\}$ with $m$ an integer.
Continuing from Eq.~(\ref{eq: generating function integral form}) and substituting $\mathbf{r} = 0$ as before we get for the first set of images,
\begin{equation}
    \tilde{W}_{z,1}^*(0) = \frac{1}{(2 \pi)^2} \iint \frac{d k_x d k_y}{1-z \lambda(\mathbf{k})} \sum_{m=-\infty}^{\infty} e^{-i k_y 2 m w}.
\end{equation}
Using the technique from Sec.~\ref{SM: one}, we use the delta function representation from Eq.~\eqref{eq: delta function representation},
\begin{equation}
    \tilde{W}_{z,1}^*(0) = \frac{1}{4 \pi w} \sum_{m=0}^{2 w-1} \int_{-\pi}^\pi \frac{d k_x}{1-z \lambda\left(\mathbf{k}=\left(k x, \frac{n \pi}{w}\right)\right)}.
\end{equation}
The integral appearing inside the sum was evaluated in section \ref{SM: one}, we now have
\begin{eqnarray}
    &&\tilde{W}_{z,1}^*(0) = \\ \nonumber
    && \frac{1}{2 w} \sum_{m=0}^{2 w-1} \int_0^{\infty} d s e^{-s} I_0\left(2 s z B\right) \exp \left[2 s z B \cos \left(\frac{m \pi}{w}\right)\right].
\end{eqnarray}
For the second set of image charges we have the same solution but with a factor $e^{- i k_y w}$ which after performing the integral over $k_y$ turns into $e^{-i m \pi}=(-1)^m$, and therefore
\begin{eqnarray}
    &&\tilde{W}_{z,2}^*(0) = \frac{1}{2 w} \sum_{m^{\prime}=0}^{2 w-1} (-1)^{m^{\prime}} \\ \nonumber
    &\times&  \int_0^{\infty} d s e^{-s} I_0\left(2 s z B\right) \exp \left[2 s z B \cos \left(\frac{ m^{\prime} \pi}{w}\right)\right].
\end{eqnarray}
The final result is 
\begin{eqnarray}
    && \tilde{W}_{z}^*(0) = \tilde{W}_{z,1}^*(0)+\tilde{W}_{z,2}^*(0) \\ \nonumber
    &&= \frac{1}{w} \sum_{m=0}^{w-1} \int d s e^{-s} I_0\left(2 s z B\right) \exp \left[2 s z B \cos \left(\frac{2 m \pi}{w}\right)\right].
\end{eqnarray}
Which gives the same return probability as in the periodic boundary conditions.

\subsection{The mobility enhancement as a function of time \label{SM: five}}
\begin{figure}[ht!]
\centering
\includegraphics[width=0.4\textwidth]{"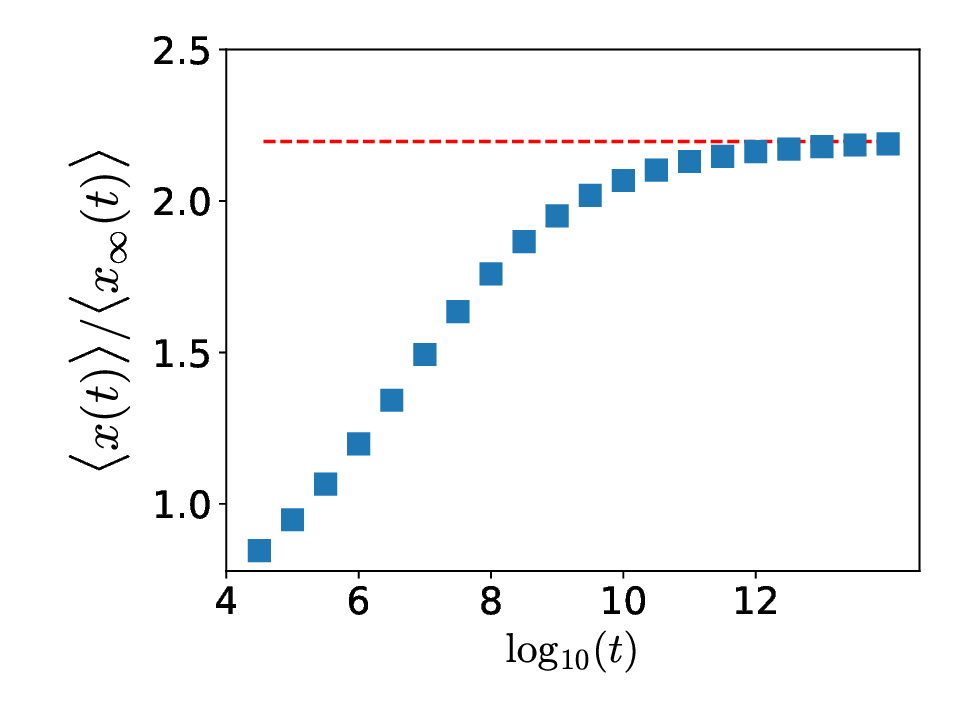"}
\caption{Enhancement of mobility in a channel of width $w=5$ ($\langle x(t)\rangle$) with respect to mobility in unrestricted $2$d geometry ($\langle x_\infty(t)\rangle$), as a function of time. Constants are $F=0.1$, $A=1$ and $\alpha=0.3$. The red dashed line is our theoretical description in Eq.~\eqref{eq: mobility enhancement exact} valid for large times. Simulations are in symbols (squares) and represent a disorder average over $300$ fixed energetic landscapes with $10^4$ trajectories each.
}
\label{Fig:time}
\end{figure}
In this section we investigate the channel's mobility enhancement effect as a function of time using simulations.
While our theoretical description of the mobility enhancement (see Eq.~(10) in the main text) applies only for $t\to\infty$ we ran simulations to see the behavior for smaller times.
We see in Fig.\ref{Fig:time} that the effect is also preserved for smaller times.


\bibliography{./main.bib}

\begin{thebibliography}{78}%
\makeatletter
\providecommand \@ifxundefined [1]{%
 \@ifx{#1\undefined}
}%
\providecommand \@ifnum [1]{%
 \ifnum #1\expandafter \@firstoftwo
 \else \expandafter \@secondoftwo
 \fi
}%
\providecommand \@ifx [1]{%
 \ifx #1\expandafter \@firstoftwo
 \else \expandafter \@secondoftwo
 \fi
}%
\providecommand \natexlab [1]{#1}%
\providecommand \enquote  [1]{``#1''}%
\providecommand \bibnamefont  [1]{#1}%
\providecommand \bibfnamefont [1]{#1}%
\providecommand \citenamefont [1]{#1}%
\providecommand \href@noop [0]{\@secondoftwo}%
\providecommand \href [0]{\begingroup \@sanitize@url \@href}%
\providecommand \@href[1]{\@@startlink{#1}\@@href}%
\providecommand \@@href[1]{\endgroup#1\@@endlink}%
\providecommand \@sanitize@url [0]{\catcode `\\12\catcode `\$12\catcode `\&12\catcode `\#12\catcode `\^12\catcode `\_12\catcode `\%12\relax}%
\providecommand \@@startlink[1]{}%
\providecommand \@@endlink[0]{}%
\providecommand \url  [0]{\begingroup\@sanitize@url \@url }%
\providecommand \@url [1]{\endgroup\@href {#1}{\urlprefix }}%
\providecommand \urlprefix  [0]{URL }%
\providecommand \Eprint [0]{\href }%
\providecommand \doibase [0]{https://doi.org/}%
\providecommand \selectlanguage [0]{\@gobble}%
\providecommand \bibinfo  [0]{\@secondoftwo}%
\providecommand \bibfield  [0]{\@secondoftwo}%
\providecommand \translation [1]{[#1]}%
\providecommand \BibitemOpen [0]{}%
\providecommand \bibitemStop [0]{}%
\providecommand \bibitemNoStop [0]{.\EOS\space}%
\providecommand \EOS [0]{\spacefactor3000\relax}%
\providecommand \BibitemShut  [1]{\csname bibitem#1\endcsname}%
\let\auto@bib@innerbib\@empty
\bibitem [{\citenamefont {Scott}\ \emph {et~al.}(2023)\citenamefont {Scott}, \citenamefont {Weiss}, \citenamefont {Selhuber-Unkel}, \citenamefont {Barooji}, \citenamefont {Sabri}, \citenamefont {Erler}, \citenamefont {Metzler},\ and\ \citenamefont {Oddershede}}]{scott2023extracting}%
  \BibitemOpen
  \bibfield  {author} {\bibinfo {author} {\bibfnamefont {S.}~\bibnamefont {Scott}}, \bibinfo {author} {\bibfnamefont {M.}~\bibnamefont {Weiss}}, \bibinfo {author} {\bibfnamefont {C.}~\bibnamefont {Selhuber-Unkel}}, \bibinfo {author} {\bibfnamefont {Y.~F.}\ \bibnamefont {Barooji}}, \bibinfo {author} {\bibfnamefont {A.}~\bibnamefont {Sabri}}, \bibinfo {author} {\bibfnamefont {J.~T.}\ \bibnamefont {Erler}}, \bibinfo {author} {\bibfnamefont {R.}~\bibnamefont {Metzler}},\ and\ \bibinfo {author} {\bibfnamefont {L.~B.}\ \bibnamefont {Oddershede}},\ }\bibfield  {title} {\bibinfo {title} {Extracting, quantifying, and comparing dynamical and biomechanical properties of living matter through single particle tracking},\ }\href@noop {} {\bibfield  {journal} {\bibinfo  {journal} {Physical Chemistry Chemical Physics}\ }\textbf {\bibinfo {volume} {25}},\ \bibinfo {pages} {1513} (\bibinfo {year} {2023})}\BibitemShut {NoStop}%
\bibitem [{\citenamefont {Waigh}\ and\ \citenamefont {Korabel}(2023)}]{waigh2023heterogeneous}%
  \BibitemOpen
  \bibfield  {author} {\bibinfo {author} {\bibfnamefont {T.~A.}\ \bibnamefont {Waigh}}\ and\ \bibinfo {author} {\bibfnamefont {N.}~\bibnamefont {Korabel}},\ }\bibfield  {title} {\bibinfo {title} {Heterogeneous anomalous transport in cellular and molecular biology},\ }\href@noop {} {\bibfield  {journal} {\bibinfo  {journal} {Reports on Progress in Physics}\ } (\bibinfo {year} {2023})}\BibitemShut {NoStop}%
\bibitem [{\citenamefont {Bouchaud}\ and\ \citenamefont {Georges}(1990)}]{bouchaud1990anomalous}%
  \BibitemOpen
  \bibfield  {author} {\bibinfo {author} {\bibfnamefont {J.-P.}\ \bibnamefont {Bouchaud}}\ and\ \bibinfo {author} {\bibfnamefont {A.}~\bibnamefont {Georges}},\ }\bibfield  {title} {\bibinfo {title} {Anomalous diffusion in disordered media: statistical mechanisms, models and physical applications},\ }\href@noop {} {\bibfield  {journal} {\bibinfo  {journal} {Physics reports}\ }\textbf {\bibinfo {volume} {195}},\ \bibinfo {pages} {127} (\bibinfo {year} {1990})}\BibitemShut {NoStop}%
\bibitem [{\citenamefont {Metzler}\ and\ \citenamefont {Klafter}(2000)}]{metzler2000random}%
  \BibitemOpen
  \bibfield  {author} {\bibinfo {author} {\bibfnamefont {R.}~\bibnamefont {Metzler}}\ and\ \bibinfo {author} {\bibfnamefont {J.}~\bibnamefont {Klafter}},\ }\bibfield  {title} {\bibinfo {title} {The random walk's guide to anomalous diffusion: a fractional dynamics approach},\ }\href@noop {} {\bibfield  {journal} {\bibinfo  {journal} {Physics reports}\ }\textbf {\bibinfo {volume} {339}},\ \bibinfo {pages} {1} (\bibinfo {year} {2000})}\BibitemShut {NoStop}%
\bibitem [{\citenamefont {Montroll}\ and\ \citenamefont {West}(1979)}]{montroll1979enriched}%
  \BibitemOpen
  \bibfield  {author} {\bibinfo {author} {\bibfnamefont {E.~W.}\ \bibnamefont {Montroll}}\ and\ \bibinfo {author} {\bibfnamefont {B.~J.}\ \bibnamefont {West}},\ }\bibfield  {title} {\bibinfo {title} {On an enriched collection of stochastic processes},\ }\href@noop {} {\bibfield  {journal} {\bibinfo  {journal} {Fluctuation phenomena}\ }\textbf {\bibinfo {volume} {66}},\ \bibinfo {pages} {61} (\bibinfo {year} {1979})}\BibitemShut {NoStop}%
\bibitem [{\citenamefont {Metzler}\ \emph {et~al.}(2014)\citenamefont {Metzler}, \citenamefont {Jeon}, \citenamefont {Cherstvy},\ and\ \citenamefont {Barkai}}]{metzler2014anomalous}%
  \BibitemOpen
  \bibfield  {author} {\bibinfo {author} {\bibfnamefont {R.}~\bibnamefont {Metzler}}, \bibinfo {author} {\bibfnamefont {J.-H.}\ \bibnamefont {Jeon}}, \bibinfo {author} {\bibfnamefont {A.~G.}\ \bibnamefont {Cherstvy}},\ and\ \bibinfo {author} {\bibfnamefont {E.}~\bibnamefont {Barkai}},\ }\bibfield  {title} {\bibinfo {title} {Anomalous diffusion models and their properties: non-stationarity, non-ergodicity, and ageing at the centenary of single particle tracking},\ }\href@noop {} {\bibfield  {journal} {\bibinfo  {journal} {Physical Chemistry Chemical Physics}\ }\textbf {\bibinfo {volume} {16}},\ \bibinfo {pages} {24128} (\bibinfo {year} {2014})}\BibitemShut {NoStop}%
\bibitem [{\citenamefont {Barkai}\ \emph {et~al.}(2012{\natexlab{a}})\citenamefont {Barkai}, \citenamefont {Garini},\ and\ \citenamefont {Metzler}}]{barkai2012single}%
  \BibitemOpen
  \bibfield  {author} {\bibinfo {author} {\bibfnamefont {E.}~\bibnamefont {Barkai}}, \bibinfo {author} {\bibfnamefont {Y.}~\bibnamefont {Garini}},\ and\ \bibinfo {author} {\bibfnamefont {R.}~\bibnamefont {Metzler}},\ }\bibfield  {title} {\bibinfo {title} {of single molecules in living cells},\ }\href@noop {} {\bibfield  {journal} {\bibinfo  {journal} {Physics Today}\ }\textbf {\bibinfo {volume} {65}},\ \bibinfo {pages} {29} (\bibinfo {year} {2012}{\natexlab{a}})}\BibitemShut {NoStop}%
\bibitem [{\citenamefont {Anderson}\ \emph {et~al.}(2019)\citenamefont {Anderson}, \citenamefont {Matsuda}, \citenamefont {Garamella}, \citenamefont {Peddireddy}, \citenamefont {Robertson-Anderson},\ and\ \citenamefont {McGorty}}]{anderson2019filament}%
  \BibitemOpen
  \bibfield  {author} {\bibinfo {author} {\bibfnamefont {S.~J.}\ \bibnamefont {Anderson}}, \bibinfo {author} {\bibfnamefont {C.}~\bibnamefont {Matsuda}}, \bibinfo {author} {\bibfnamefont {J.}~\bibnamefont {Garamella}}, \bibinfo {author} {\bibfnamefont {K.~R.}\ \bibnamefont {Peddireddy}}, \bibinfo {author} {\bibfnamefont {R.~M.}\ \bibnamefont {Robertson-Anderson}},\ and\ \bibinfo {author} {\bibfnamefont {R.}~\bibnamefont {McGorty}},\ }\bibfield  {title} {\bibinfo {title} {Filament rigidity vies with mesh size in determining anomalous diffusion in cytoskeleton},\ }\href@noop {} {\bibfield  {journal} {\bibinfo  {journal} {Biomacromolecules}\ }\textbf {\bibinfo {volume} {20}},\ \bibinfo {pages} {4380} (\bibinfo {year} {2019})}\BibitemShut {NoStop}%
\bibitem [{\citenamefont {Yamamoto}\ \emph {et~al.}(2021)\citenamefont {Yamamoto}, \citenamefont {Akimoto}, \citenamefont {Mitsutake},\ and\ \citenamefont {Metzler}}]{yamamoto2021universal}%
  \BibitemOpen
  \bibfield  {author} {\bibinfo {author} {\bibfnamefont {E.}~\bibnamefont {Yamamoto}}, \bibinfo {author} {\bibfnamefont {T.}~\bibnamefont {Akimoto}}, \bibinfo {author} {\bibfnamefont {A.}~\bibnamefont {Mitsutake}},\ and\ \bibinfo {author} {\bibfnamefont {R.}~\bibnamefont {Metzler}},\ }\bibfield  {title} {\bibinfo {title} {Universal relation between instantaneous diffusivity and radius of gyration of proteins in aqueous solution},\ }\href@noop {} {\bibfield  {journal} {\bibinfo  {journal} {Physical review letters}\ }\textbf {\bibinfo {volume} {126}},\ \bibinfo {pages} {128101} (\bibinfo {year} {2021})}\BibitemShut {NoStop}%
\bibitem [{\citenamefont {Metzler}\ \emph {et~al.}(2022)\citenamefont {Metzler}, \citenamefont {Rajyaguru},\ and\ \citenamefont {Berkowitz}}]{metzler2022modelling}%
  \BibitemOpen
  \bibfield  {author} {\bibinfo {author} {\bibfnamefont {R.}~\bibnamefont {Metzler}}, \bibinfo {author} {\bibfnamefont {A.}~\bibnamefont {Rajyaguru}},\ and\ \bibinfo {author} {\bibfnamefont {B.}~\bibnamefont {Berkowitz}},\ }\bibfield  {title} {\bibinfo {title} {Modelling anomalous diffusion in semi-infinite disordered systems and porous media},\ }\href@noop {} {\bibfield  {journal} {\bibinfo  {journal} {New journal of physics}\ }\textbf {\bibinfo {volume} {24}},\ \bibinfo {pages} {123004} (\bibinfo {year} {2022})}\BibitemShut {NoStop}%
\bibitem [{\citenamefont {Malgaretti}\ \emph {et~al.}(2019)\citenamefont {Malgaretti}, \citenamefont {Oshanin},\ and\ \citenamefont {Talbot}}]{malgaretti2019special}%
  \BibitemOpen
  \bibfield  {author} {\bibinfo {author} {\bibfnamefont {P.}~\bibnamefont {Malgaretti}}, \bibinfo {author} {\bibfnamefont {G.}~\bibnamefont {Oshanin}},\ and\ \bibinfo {author} {\bibfnamefont {J.}~\bibnamefont {Talbot}},\ }\bibfield  {title} {\bibinfo {title} {Special issue on transport in narrow channels},\ }\href@noop {} {\bibfield  {journal} {\bibinfo  {journal} {Journal of Physics: Condensed Matter}\ }\textbf {\bibinfo {volume} {31}},\ \bibinfo {pages} {270201} (\bibinfo {year} {2019})}\BibitemShut {NoStop}%
\bibitem [{\citenamefont {Kondrat}\ \emph {et~al.}(2014{\natexlab{a}})\citenamefont {Kondrat}, \citenamefont {Wu}, \citenamefont {Qiao},\ and\ \citenamefont {Kornyshev}}]{kondrat2014accelerating}%
  \BibitemOpen
  \bibfield  {author} {\bibinfo {author} {\bibfnamefont {S.}~\bibnamefont {Kondrat}}, \bibinfo {author} {\bibfnamefont {P.}~\bibnamefont {Wu}}, \bibinfo {author} {\bibfnamefont {R.}~\bibnamefont {Qiao}},\ and\ \bibinfo {author} {\bibfnamefont {A.~A.}\ \bibnamefont {Kornyshev}},\ }\bibfield  {title} {\bibinfo {title} {Accelerating charging dynamics in subnanometre pores},\ }\href@noop {} {\bibfield  {journal} {\bibinfo  {journal} {Nature materials}\ }\textbf {\bibinfo {volume} {13}},\ \bibinfo {pages} {387} (\bibinfo {year} {2014}{\natexlab{a}})}\BibitemShut {NoStop}%
\bibitem [{\citenamefont {Kondrat}\ \emph {et~al.}(2014{\natexlab{b}})\citenamefont {Kondrat}, \citenamefont {Oshanin}, \citenamefont {Kornyshev} \emph {et~al.}}]{kondrat2014charging}%
  \BibitemOpen
  \bibfield  {author} {\bibinfo {author} {\bibfnamefont {S.}~\bibnamefont {Kondrat}}, \bibinfo {author} {\bibfnamefont {G.}~\bibnamefont {Oshanin}}, \bibinfo {author} {\bibfnamefont {A.~A.}\ \bibnamefont {Kornyshev}}, \emph {et~al.},\ }\bibfield  {title} {\bibinfo {title} {Charging dynamics of supercapacitors with narrow cylindrical nanopores},\ }\href@noop {} {\bibfield  {journal} {\bibinfo  {journal} {Nanotechnology}\ }\textbf {\bibinfo {volume} {25}},\ \bibinfo {pages} {315401} (\bibinfo {year} {2014}{\natexlab{b}})}\BibitemShut {NoStop}%
\bibitem [{\citenamefont {B{\'e}nichou}\ \emph {et~al.}(2018)\citenamefont {B{\'e}nichou}, \citenamefont {Illien}, \citenamefont {Oshanin}, \citenamefont {Sarracino},\ and\ \citenamefont {Voituriez}}]{benichou2018tracer}%
  \BibitemOpen
  \bibfield  {author} {\bibinfo {author} {\bibfnamefont {O.}~\bibnamefont {B{\'e}nichou}}, \bibinfo {author} {\bibfnamefont {P.}~\bibnamefont {Illien}}, \bibinfo {author} {\bibfnamefont {G.}~\bibnamefont {Oshanin}}, \bibinfo {author} {\bibfnamefont {A.}~\bibnamefont {Sarracino}},\ and\ \bibinfo {author} {\bibfnamefont {R.}~\bibnamefont {Voituriez}},\ }\bibfield  {title} {\bibinfo {title} {Tracer diffusion in crowded narrow channels},\ }\href@noop {} {\bibfield  {journal} {\bibinfo  {journal} {Journal of Physics: Condensed Matter}\ }\textbf {\bibinfo {volume} {30}},\ \bibinfo {pages} {443001} (\bibinfo {year} {2018})}\BibitemShut {NoStop}%
\bibitem [{\citenamefont {Uspal}\ \emph {et~al.}(2013)\citenamefont {Uspal}, \citenamefont {Burak~Eral},\ and\ \citenamefont {Doyle}}]{uspal2013engineering}%
  \BibitemOpen
  \bibfield  {author} {\bibinfo {author} {\bibfnamefont {W.~E.}\ \bibnamefont {Uspal}}, \bibinfo {author} {\bibfnamefont {H.}~\bibnamefont {Burak~Eral}},\ and\ \bibinfo {author} {\bibfnamefont {P.~S.}\ \bibnamefont {Doyle}},\ }\bibfield  {title} {\bibinfo {title} {Engineering particle trajectories in microfluidic flows using particle shape},\ }\href@noop {} {\bibfield  {journal} {\bibinfo  {journal} {Nature communications}\ }\textbf {\bibinfo {volume} {4}},\ \bibinfo {pages} {2666} (\bibinfo {year} {2013})}\BibitemShut {NoStop}%
\bibitem [{\citenamefont {Sarracino}\ \emph {et~al.}(2016)\citenamefont {Sarracino}, \citenamefont {Cecconi}, \citenamefont {Puglisi},\ and\ \citenamefont {Vulpiani}}]{sarracino2016nonlinear}%
  \BibitemOpen
  \bibfield  {author} {\bibinfo {author} {\bibfnamefont {A.}~\bibnamefont {Sarracino}}, \bibinfo {author} {\bibfnamefont {F.}~\bibnamefont {Cecconi}}, \bibinfo {author} {\bibfnamefont {A.}~\bibnamefont {Puglisi}},\ and\ \bibinfo {author} {\bibfnamefont {A.}~\bibnamefont {Vulpiani}},\ }\bibfield  {title} {\bibinfo {title} {Nonlinear response of inertial tracers in steady laminar flows: Differential and absolute negative mobility},\ }\href@noop {} {\bibfield  {journal} {\bibinfo  {journal} {Physical review letters}\ }\textbf {\bibinfo {volume} {117}},\ \bibinfo {pages} {174501} (\bibinfo {year} {2016})}\BibitemShut {NoStop}%
\bibitem [{\citenamefont {Cecconi}\ \emph {et~al.}(2017)\citenamefont {Cecconi}, \citenamefont {Puglisi}, \citenamefont {Sarracino},\ and\ \citenamefont {Vulpiani}}]{cecconi2017anomalous}%
  \BibitemOpen
  \bibfield  {author} {\bibinfo {author} {\bibfnamefont {F.}~\bibnamefont {Cecconi}}, \bibinfo {author} {\bibfnamefont {A.}~\bibnamefont {Puglisi}}, \bibinfo {author} {\bibfnamefont {A.}~\bibnamefont {Sarracino}},\ and\ \bibinfo {author} {\bibfnamefont {A.}~\bibnamefont {Vulpiani}},\ }\bibfield  {title} {\bibinfo {title} {Anomalous force-velocity relation of driven inertial tracers in steady laminar flows},\ }\href@noop {} {\bibfield  {journal} {\bibinfo  {journal} {The European Physical Journal E}\ }\textbf {\bibinfo {volume} {40}},\ \bibinfo {pages} {1} (\bibinfo {year} {2017})}\BibitemShut {NoStop}%
\bibitem [{\citenamefont {Cecconi}\ \emph {et~al.}(2018)\citenamefont {Cecconi}, \citenamefont {Puglisi}, \citenamefont {Sarracino},\ and\ \citenamefont {Vulpiani}}]{cecconi2018anomalous}%
  \BibitemOpen
  \bibfield  {author} {\bibinfo {author} {\bibfnamefont {F.}~\bibnamefont {Cecconi}}, \bibinfo {author} {\bibfnamefont {A.}~\bibnamefont {Puglisi}}, \bibinfo {author} {\bibfnamefont {A.}~\bibnamefont {Sarracino}},\ and\ \bibinfo {author} {\bibfnamefont {A.}~\bibnamefont {Vulpiani}},\ }\bibfield  {title} {\bibinfo {title} {Anomalous mobility of a driven active particle in a steady laminar flow},\ }\href@noop {} {\bibfield  {journal} {\bibinfo  {journal} {Journal of Physics: Condensed Matter}\ }\textbf {\bibinfo {volume} {30}},\ \bibinfo {pages} {264002} (\bibinfo {year} {2018})}\BibitemShut {NoStop}%
\bibitem [{\citenamefont {Ashcroft}\ and\ \citenamefont {Mermin}(1976)}]{ashcroft1976solid}%
  \BibitemOpen
  \bibfield  {author} {\bibinfo {author} {\bibfnamefont {N.~W.}\ \bibnamefont {Ashcroft}}\ and\ \bibinfo {author} {\bibfnamefont {N.}~\bibnamefont {Mermin}},\ }\bibfield  {title} {\bibinfo {title} {Solid state},\ }\href@noop {} {\bibfield  {journal} {\bibinfo  {journal} {Physics (New York: Holt, Rinehart and Winston)}\ } (\bibinfo {year} {1976})}\BibitemShut {NoStop}%
\bibitem [{\citenamefont {Scharfe}(1970)}]{scharfe1970transient}%
  \BibitemOpen
  \bibfield  {author} {\bibinfo {author} {\bibfnamefont {M.}~\bibnamefont {Scharfe}},\ }\bibfield  {title} {\bibinfo {title} {Transient photoconductivity in vitreous as 2 se 3},\ }\href@noop {} {\bibfield  {journal} {\bibinfo  {journal} {Physical Review B}\ }\textbf {\bibinfo {volume} {2}},\ \bibinfo {pages} {5025} (\bibinfo {year} {1970})}\BibitemShut {NoStop}%
\bibitem [{\citenamefont {Pai}\ and\ \citenamefont {Scharfe}(1972)}]{pai1972charge}%
  \BibitemOpen
  \bibfield  {author} {\bibinfo {author} {\bibfnamefont {D.~M.}\ \bibnamefont {Pai}}\ and\ \bibinfo {author} {\bibfnamefont {M.~E.}\ \bibnamefont {Scharfe}},\ }\bibfield  {title} {\bibinfo {title} {Charge transport in films of amorphous arsenic triselenide},\ }\href@noop {} {\bibfield  {journal} {\bibinfo  {journal} {Journal of Non-Crystalline Solids}\ }\textbf {\bibinfo {volume} {8}},\ \bibinfo {pages} {752} (\bibinfo {year} {1972})}\BibitemShut {NoStop}%
\bibitem [{\citenamefont {Pfister}(1974)}]{pfister1974pressure}%
  \BibitemOpen
  \bibfield  {author} {\bibinfo {author} {\bibfnamefont {G.}~\bibnamefont {Pfister}},\ }\bibfield  {title} {\bibinfo {title} {Pressure-dependent electronic transport in amorphous as 2 se 3},\ }\href@noop {} {\bibfield  {journal} {\bibinfo  {journal} {Physical Review Letters}\ }\textbf {\bibinfo {volume} {33}},\ \bibinfo {pages} {1474} (\bibinfo {year} {1974})}\BibitemShut {NoStop}%
\bibitem [{\citenamefont {Mort}\ and\ \citenamefont {Lakatos}(1970)}]{mort1970steady}%
  \BibitemOpen
  \bibfield  {author} {\bibinfo {author} {\bibfnamefont {J.}~\bibnamefont {Mort}}\ and\ \bibinfo {author} {\bibfnamefont {A.}~\bibnamefont {Lakatos}},\ }\bibfield  {title} {\bibinfo {title} {Steady state and transient photoemission into amorphous insulators},\ }\href@noop {} {\bibfield  {journal} {\bibinfo  {journal} {Journal of Non-Crystalline Solids}\ }\textbf {\bibinfo {volume} {4}},\ \bibinfo {pages} {117} (\bibinfo {year} {1970})}\BibitemShut {NoStop}%
\bibitem [{\citenamefont {Gill}(1972)}]{gill1972drift}%
  \BibitemOpen
  \bibfield  {author} {\bibinfo {author} {\bibfnamefont {W.}~\bibnamefont {Gill}},\ }\bibfield  {title} {\bibinfo {title} {Drift mobilities in amorphous charge-transfer complexes of trinitrofluorenone and poly-n-vinylcarbazole},\ }\href@noop {} {\bibfield  {journal} {\bibinfo  {journal} {Journal of Applied Physics}\ }\textbf {\bibinfo {volume} {43}},\ \bibinfo {pages} {5033} (\bibinfo {year} {1972})}\BibitemShut {NoStop}%
\bibitem [{\citenamefont {Scher}\ and\ \citenamefont {Montroll}(1975)}]{scher1975anomalous}%
  \BibitemOpen
  \bibfield  {author} {\bibinfo {author} {\bibfnamefont {H.}~\bibnamefont {Scher}}\ and\ \bibinfo {author} {\bibfnamefont {E.~W.}\ \bibnamefont {Montroll}},\ }\bibfield  {title} {\bibinfo {title} {Anomalous transit-time dispersion in amorphous solids},\ }\href@noop {} {\bibfield  {journal} {\bibinfo  {journal} {Physical Review B}\ }\textbf {\bibinfo {volume} {12}},\ \bibinfo {pages} {2455} (\bibinfo {year} {1975})}\BibitemShut {NoStop}%
\bibitem [{\citenamefont {Yuan}\ \emph {et~al.}(2000)\citenamefont {Yuan}, \citenamefont {Gregg},\ and\ \citenamefont {Lawrence}}]{yuan2000time}%
  \BibitemOpen
  \bibfield  {author} {\bibinfo {author} {\bibfnamefont {Y.}~\bibnamefont {Yuan}}, \bibinfo {author} {\bibfnamefont {B.~A.}\ \bibnamefont {Gregg}},\ and\ \bibinfo {author} {\bibfnamefont {M.~F.}\ \bibnamefont {Lawrence}},\ }\bibfield  {title} {\bibinfo {title} {Time-of-flight study of electrical charge mobilities in liquid-crystalline zinc octakis ($\beta$-octoxyethyl) porphyrin films},\ }\href@noop {} {\bibfield  {journal} {\bibinfo  {journal} {Journal of Materials Research}\ }\textbf {\bibinfo {volume} {15}},\ \bibinfo {pages} {2494} (\bibinfo {year} {2000})}\BibitemShut {NoStop}%
\bibitem [{\citenamefont {Tyutnev}\ \emph {et~al.}(2017)\citenamefont {Tyutnev}, \citenamefont {Weiss}, \citenamefont {Saenko},\ and\ \citenamefont {Pozhidaev}}]{tyutnev2017mobility}%
  \BibitemOpen
  \bibfield  {author} {\bibinfo {author} {\bibfnamefont {A.}~\bibnamefont {Tyutnev}}, \bibinfo {author} {\bibfnamefont {D.}~\bibnamefont {Weiss}}, \bibinfo {author} {\bibfnamefont {V.}~\bibnamefont {Saenko}},\ and\ \bibinfo {author} {\bibfnamefont {E.}~\bibnamefont {Pozhidaev}},\ }\bibfield  {title} {\bibinfo {title} {About mobility thickness dependence in molecularly doped polymers},\ }\href@noop {} {\bibfield  {journal} {\bibinfo  {journal} {Chemical Physics}\ }\textbf {\bibinfo {volume} {495}},\ \bibinfo {pages} {16} (\bibinfo {year} {2017})}\BibitemShut {NoStop}%
\bibitem [{\citenamefont {Dunlap}(1996)}]{dunlap1996hopping}%
  \BibitemOpen
  \bibfield  {author} {\bibinfo {author} {\bibfnamefont {D.}~\bibnamefont {Dunlap}},\ }\bibfield  {title} {\bibinfo {title} {Hopping transport in molecularly doped polymers: On the relation between disorder and a field-dependent mobility},\ }\href@noop {} {\bibfield  {journal} {\bibinfo  {journal} {Journal of Imaging Science and Technology}\ }\textbf {\bibinfo {volume} {40}},\ \bibinfo {pages} {291} (\bibinfo {year} {1996})}\BibitemShut {NoStop}%
\bibitem [{\citenamefont {Scholz}\ \emph {et~al.}(2016)\citenamefont {Scholz}, \citenamefont {Burov}, \citenamefont {Weirich}, \citenamefont {Scholz}, \citenamefont {Tabei}, \citenamefont {Gardel},\ and\ \citenamefont {Dinner}}]{scholz2016cycling}%
  \BibitemOpen
  \bibfield  {author} {\bibinfo {author} {\bibfnamefont {M.}~\bibnamefont {Scholz}}, \bibinfo {author} {\bibfnamefont {S.}~\bibnamefont {Burov}}, \bibinfo {author} {\bibfnamefont {K.~L.}\ \bibnamefont {Weirich}}, \bibinfo {author} {\bibfnamefont {B.~J.}\ \bibnamefont {Scholz}}, \bibinfo {author} {\bibfnamefont {S.~A.}\ \bibnamefont {Tabei}}, \bibinfo {author} {\bibfnamefont {M.~L.}\ \bibnamefont {Gardel}},\ and\ \bibinfo {author} {\bibfnamefont {A.~R.}\ \bibnamefont {Dinner}},\ }\bibfield  {title} {\bibinfo {title} {Cycling state that can lead to glassy dynamics in intracellular transport},\ }\href@noop {} {\bibfield  {journal} {\bibinfo  {journal} {Physical Review X}\ }\textbf {\bibinfo {volume} {6}},\ \bibinfo {pages} {011037} (\bibinfo {year} {2016})}\BibitemShut {NoStop}%
\bibitem [{\citenamefont {Barkai}\ \emph {et~al.}(2012{\natexlab{b}})\citenamefont {Barkai}, \citenamefont {Garini},\ and\ \citenamefont {Metzler}}]{barkai2012strange}%
  \BibitemOpen
  \bibfield  {author} {\bibinfo {author} {\bibfnamefont {E.}~\bibnamefont {Barkai}}, \bibinfo {author} {\bibfnamefont {Y.}~\bibnamefont {Garini}},\ and\ \bibinfo {author} {\bibfnamefont {R.}~\bibnamefont {Metzler}},\ }\bibfield  {title} {\bibinfo {title} {Strange kinetics of single molecules in living cells},\ }\href@noop {} {\bibfield  {journal} {\bibinfo  {journal} {Physics today}\ }\textbf {\bibinfo {volume} {65}},\ \bibinfo {pages} {29} (\bibinfo {year} {2012}{\natexlab{b}})}\BibitemShut {NoStop}%
\bibitem [{\citenamefont {Kompella}\ \emph {et~al.}(2024)\citenamefont {Kompella}, \citenamefont {Romano}, \citenamefont {Stansfield},\ and\ \citenamefont {Mancera}}]{kompella2024determines}%
  \BibitemOpen
  \bibfield  {author} {\bibinfo {author} {\bibfnamefont {V.~P.~S.}\ \bibnamefont {Kompella}}, \bibinfo {author} {\bibfnamefont {M.~C.}\ \bibnamefont {Romano}}, \bibinfo {author} {\bibfnamefont {I.}~\bibnamefont {Stansfield}},\ and\ \bibinfo {author} {\bibfnamefont {R.~L.}\ \bibnamefont {Mancera}},\ }\bibfield  {title} {\bibinfo {title} {What determines sub-diffusive behavior in crowded protein solutions?},\ }\href@noop {} {\bibfield  {journal} {\bibinfo  {journal} {Biophysical Journal}\ }\textbf {\bibinfo {volume} {123}},\ \bibinfo {pages} {134} (\bibinfo {year} {2024})}\BibitemShut {NoStop}%
\bibitem [{\citenamefont {Yu}\ \emph {et~al.}(2018)\citenamefont {Yu}, \citenamefont {Sheats}, \citenamefont {Cicuta}, \citenamefont {Sclavi}, \citenamefont {Cosentino~Lagomarsino},\ and\ \citenamefont {Dorfman}}]{yu2018subdiffusion}%
  \BibitemOpen
  \bibfield  {author} {\bibinfo {author} {\bibfnamefont {S.}~\bibnamefont {Yu}}, \bibinfo {author} {\bibfnamefont {J.}~\bibnamefont {Sheats}}, \bibinfo {author} {\bibfnamefont {P.}~\bibnamefont {Cicuta}}, \bibinfo {author} {\bibfnamefont {B.}~\bibnamefont {Sclavi}}, \bibinfo {author} {\bibfnamefont {M.}~\bibnamefont {Cosentino~Lagomarsino}},\ and\ \bibinfo {author} {\bibfnamefont {K.~D.}\ \bibnamefont {Dorfman}},\ }\bibfield  {title} {\bibinfo {title} {Subdiffusion of loci and cytoplasmic particles are different in compressed escherichia coli cells},\ }\href@noop {} {\bibfield  {journal} {\bibinfo  {journal} {Communications biology}\ }\textbf {\bibinfo {volume} {1}},\ \bibinfo {pages} {176} (\bibinfo {year} {2018})}\BibitemShut {NoStop}%
\bibitem [{\citenamefont {Sabri}\ \emph {et~al.}(2020)\citenamefont {Sabri}, \citenamefont {Xu}, \citenamefont {Krapf},\ and\ \citenamefont {Weiss}}]{sabri2020elucidating}%
  \BibitemOpen
  \bibfield  {author} {\bibinfo {author} {\bibfnamefont {A.}~\bibnamefont {Sabri}}, \bibinfo {author} {\bibfnamefont {X.}~\bibnamefont {Xu}}, \bibinfo {author} {\bibfnamefont {D.}~\bibnamefont {Krapf}},\ and\ \bibinfo {author} {\bibfnamefont {M.}~\bibnamefont {Weiss}},\ }\bibfield  {title} {\bibinfo {title} {Elucidating the origin of heterogeneous anomalous diffusion in the cytoplasm of mammalian cells},\ }\href@noop {} {\bibfield  {journal} {\bibinfo  {journal} {Physical Review Letters}\ }\textbf {\bibinfo {volume} {125}},\ \bibinfo {pages} {058101} (\bibinfo {year} {2020})}\BibitemShut {NoStop}%
\bibitem [{\citenamefont {Marty}\ and\ \citenamefont {Dauchot}(2005)}]{marty2005subdiffusion}%
  \BibitemOpen
  \bibfield  {author} {\bibinfo {author} {\bibfnamefont {G.}~\bibnamefont {Marty}}\ and\ \bibinfo {author} {\bibfnamefont {O.}~\bibnamefont {Dauchot}},\ }\bibfield  {title} {\bibinfo {title} {Subdiffusion and cage effect in a sheared granular material},\ }\href@noop {} {\bibfield  {journal} {\bibinfo  {journal} {Physical review letters}\ }\textbf {\bibinfo {volume} {94}},\ \bibinfo {pages} {015701} (\bibinfo {year} {2005})}\BibitemShut {NoStop}%
\bibitem [{\citenamefont {Bodrova}(2024)}]{bodrova2024diffusion}%
  \BibitemOpen
  \bibfield  {author} {\bibinfo {author} {\bibfnamefont {A.~S.}\ \bibnamefont {Bodrova}},\ }\bibfield  {title} {\bibinfo {title} {Diffusion in multicomponent granular mixtures},\ }\href@noop {} {\bibfield  {journal} {\bibinfo  {journal} {Physical Review E}\ }\textbf {\bibinfo {volume} {109}},\ \bibinfo {pages} {024903} (\bibinfo {year} {2024})}\BibitemShut {NoStop}%
\bibitem [{\citenamefont {Wong}\ \emph {et~al.}(2004)\citenamefont {Wong}, \citenamefont {Gardel}, \citenamefont {Reichman}, \citenamefont {Weeks}, \citenamefont {Valentine}, \citenamefont {Bausch},\ and\ \citenamefont {Weitz}}]{wong2004anomalous}%
  \BibitemOpen
  \bibfield  {author} {\bibinfo {author} {\bibfnamefont {I.}~\bibnamefont {Wong}}, \bibinfo {author} {\bibfnamefont {M.}~\bibnamefont {Gardel}}, \bibinfo {author} {\bibfnamefont {D.}~\bibnamefont {Reichman}}, \bibinfo {author} {\bibfnamefont {E.~R.}\ \bibnamefont {Weeks}}, \bibinfo {author} {\bibfnamefont {M.}~\bibnamefont {Valentine}}, \bibinfo {author} {\bibfnamefont {A.}~\bibnamefont {Bausch}},\ and\ \bibinfo {author} {\bibfnamefont {D.~A.}\ \bibnamefont {Weitz}},\ }\bibfield  {title} {\bibinfo {title} {Anomalous diffusion probes microstructure dynamics of entangled f-actin networks},\ }\href@noop {} {\bibfield  {journal} {\bibinfo  {journal} {Physical review letters}\ }\textbf {\bibinfo {volume} {92}},\ \bibinfo {pages} {178101} (\bibinfo {year} {2004})}\BibitemShut {NoStop}%
\bibitem [{\citenamefont {Meyer}\ and\ \citenamefont {Metzler}(2024)}]{meyer2024time}%
  \BibitemOpen
  \bibfield  {author} {\bibinfo {author} {\bibfnamefont {P.~G.}\ \bibnamefont {Meyer}}\ and\ \bibinfo {author} {\bibfnamefont {R.}~\bibnamefont {Metzler}},\ }\bibfield  {title} {\bibinfo {title} {Time scales in the dynamics of political opinions and the voter model},\ }\href@noop {} {\bibfield  {journal} {\bibinfo  {journal} {New Journal of Physics}\ } (\bibinfo {year} {2024})}\BibitemShut {NoStop}%
\bibitem [{\citenamefont {McKinley}\ \emph {et~al.}(2009)\citenamefont {McKinley}, \citenamefont {Yao},\ and\ \citenamefont {Forest}}]{mckinley2009transient}%
  \BibitemOpen
  \bibfield  {author} {\bibinfo {author} {\bibfnamefont {S.~A.}\ \bibnamefont {McKinley}}, \bibinfo {author} {\bibfnamefont {L.}~\bibnamefont {Yao}},\ and\ \bibinfo {author} {\bibfnamefont {M.~G.}\ \bibnamefont {Forest}},\ }\bibfield  {title} {\bibinfo {title} {Transient anomalous diffusion of tracer particles in soft matter},\ }\href@noop {} {\bibfield  {journal} {\bibinfo  {journal} {Journal of Rheology}\ }\textbf {\bibinfo {volume} {53}},\ \bibinfo {pages} {1487} (\bibinfo {year} {2009})}\BibitemShut {NoStop}%
\bibitem [{\citenamefont {Paoluzzi}\ \emph {et~al.}(2024)\citenamefont {Paoluzzi}, \citenamefont {Levis},\ and\ \citenamefont {Pagonabarraga}}]{paoluzzi2024flocking}%
  \BibitemOpen
  \bibfield  {author} {\bibinfo {author} {\bibfnamefont {M.}~\bibnamefont {Paoluzzi}}, \bibinfo {author} {\bibfnamefont {D.}~\bibnamefont {Levis}},\ and\ \bibinfo {author} {\bibfnamefont {I.}~\bibnamefont {Pagonabarraga}},\ }\bibfield  {title} {\bibinfo {title} {From flocking to glassiness in dense disordered polar active matter},\ }\href@noop {} {\bibfield  {journal} {\bibinfo  {journal} {Communications Physics}\ }\textbf {\bibinfo {volume} {7}},\ \bibinfo {pages} {57} (\bibinfo {year} {2024})}\BibitemShut {NoStop}%
\bibitem [{\citenamefont {Tabei}\ \emph {et~al.}(2013)\citenamefont {Tabei}, \citenamefont {Burov}, \citenamefont {Kim}, \citenamefont {Kuznetsov}, \citenamefont {Huynh}, \citenamefont {Jureller}, \citenamefont {Philipson}, \citenamefont {Dinner},\ and\ \citenamefont {Scherer}}]{tabei2013intracellular}%
  \BibitemOpen
  \bibfield  {author} {\bibinfo {author} {\bibfnamefont {S.~A.}\ \bibnamefont {Tabei}}, \bibinfo {author} {\bibfnamefont {S.}~\bibnamefont {Burov}}, \bibinfo {author} {\bibfnamefont {H.~Y.}\ \bibnamefont {Kim}}, \bibinfo {author} {\bibfnamefont {A.}~\bibnamefont {Kuznetsov}}, \bibinfo {author} {\bibfnamefont {T.}~\bibnamefont {Huynh}}, \bibinfo {author} {\bibfnamefont {J.}~\bibnamefont {Jureller}}, \bibinfo {author} {\bibfnamefont {L.~H.}\ \bibnamefont {Philipson}}, \bibinfo {author} {\bibfnamefont {A.~R.}\ \bibnamefont {Dinner}},\ and\ \bibinfo {author} {\bibfnamefont {N.~F.}\ \bibnamefont {Scherer}},\ }\bibfield  {title} {\bibinfo {title} {Intracellular transport of insulin granules is a subordinated random walk},\ }\href@noop {} {\bibfield  {journal} {\bibinfo  {journal} {Proceedings of the National Academy of Sciences}\ }\textbf {\bibinfo {volume} {110}},\ \bibinfo {pages} {4911} (\bibinfo {year} {2013})}\BibitemShut {NoStop}%
\bibitem [{\citenamefont {Vilk}\ \emph {et~al.}(2022{\natexlab{a}})\citenamefont {Vilk}, \citenamefont {Orchan}, \citenamefont {Charter}, \citenamefont {Ganot}, \citenamefont {Toledo}, \citenamefont {Nathan},\ and\ \citenamefont {Assaf}}]{vilk2022ergodicity}%
  \BibitemOpen
  \bibfield  {author} {\bibinfo {author} {\bibfnamefont {O.}~\bibnamefont {Vilk}}, \bibinfo {author} {\bibfnamefont {Y.}~\bibnamefont {Orchan}}, \bibinfo {author} {\bibfnamefont {M.}~\bibnamefont {Charter}}, \bibinfo {author} {\bibfnamefont {N.}~\bibnamefont {Ganot}}, \bibinfo {author} {\bibfnamefont {S.}~\bibnamefont {Toledo}}, \bibinfo {author} {\bibfnamefont {R.}~\bibnamefont {Nathan}},\ and\ \bibinfo {author} {\bibfnamefont {M.}~\bibnamefont {Assaf}},\ }\bibfield  {title} {\bibinfo {title} {Ergodicity breaking in area-restricted search of avian predators},\ }\href@noop {} {\bibfield  {journal} {\bibinfo  {journal} {Physical Review X}\ }\textbf {\bibinfo {volume} {12}},\ \bibinfo {pages} {031005} (\bibinfo {year} {2022}{\natexlab{a}})}\BibitemShut {NoStop}%
\bibitem [{\citenamefont {Vilk}\ \emph {et~al.}(2022{\natexlab{b}})\citenamefont {Vilk}, \citenamefont {Aghion}, \citenamefont {Nathan}, \citenamefont {Toledo}, \citenamefont {Metzler},\ and\ \citenamefont {Assaf}}]{vilk2022classification}%
  \BibitemOpen
  \bibfield  {author} {\bibinfo {author} {\bibfnamefont {O.}~\bibnamefont {Vilk}}, \bibinfo {author} {\bibfnamefont {E.}~\bibnamefont {Aghion}}, \bibinfo {author} {\bibfnamefont {R.}~\bibnamefont {Nathan}}, \bibinfo {author} {\bibfnamefont {S.}~\bibnamefont {Toledo}}, \bibinfo {author} {\bibfnamefont {R.}~\bibnamefont {Metzler}},\ and\ \bibinfo {author} {\bibfnamefont {M.}~\bibnamefont {Assaf}},\ }\bibfield  {title} {\bibinfo {title} {Classification of anomalous diffusion in animal movement data using power spectral analysis},\ }\href@noop {} {\bibfield  {journal} {\bibinfo  {journal} {Journal of Physics A: Mathematical and Theoretical}\ }\textbf {\bibinfo {volume} {55}},\ \bibinfo {pages} {334004} (\bibinfo {year} {2022}{\natexlab{b}})}\BibitemShut {NoStop}%
\bibitem [{\citenamefont {Bouchaud}(1992)}]{bouchaud1992weak}%
  \BibitemOpen
  \bibfield  {author} {\bibinfo {author} {\bibfnamefont {J.-P.}\ \bibnamefont {Bouchaud}},\ }\bibfield  {title} {\bibinfo {title} {Weak ergodicity breaking and aging in disordered systems},\ }\href@noop {} {\bibfield  {journal} {\bibinfo  {journal} {Journal de Physique I}\ }\textbf {\bibinfo {volume} {2}},\ \bibinfo {pages} {1705} (\bibinfo {year} {1992})}\BibitemShut {NoStop}%
\bibitem [{\citenamefont {Monthus}\ and\ \citenamefont {Bouchaud}(1996)}]{monthus1996models}%
  \BibitemOpen
  \bibfield  {author} {\bibinfo {author} {\bibfnamefont {C.}~\bibnamefont {Monthus}}\ and\ \bibinfo {author} {\bibfnamefont {J.-P.}\ \bibnamefont {Bouchaud}},\ }\bibfield  {title} {\bibinfo {title} {Models of traps and glass phenomenology},\ }\href@noop {} {\bibfield  {journal} {\bibinfo  {journal} {Journal of Physics A: Mathematical and General}\ }\textbf {\bibinfo {volume} {29}},\ \bibinfo {pages} {3847} (\bibinfo {year} {1996})}\BibitemShut {NoStop}%
\bibitem [{\citenamefont {Rinn}\ \emph {et~al.}(2000)\citenamefont {Rinn}, \citenamefont {Maass},\ and\ \citenamefont {Bouchaud}}]{rinn2000multiple}%
  \BibitemOpen
  \bibfield  {author} {\bibinfo {author} {\bibfnamefont {B.}~\bibnamefont {Rinn}}, \bibinfo {author} {\bibfnamefont {P.}~\bibnamefont {Maass}},\ and\ \bibinfo {author} {\bibfnamefont {J.-P.}\ \bibnamefont {Bouchaud}},\ }\bibfield  {title} {\bibinfo {title} {Multiple scaling regimes in simple aging models},\ }\href@noop {} {\bibfield  {journal} {\bibinfo  {journal} {Physical Review Letters}\ }\textbf {\bibinfo {volume} {84}},\ \bibinfo {pages} {5403} (\bibinfo {year} {2000})}\BibitemShut {NoStop}%
\bibitem [{\citenamefont {Rinn}\ \emph {et~al.}(2001)\citenamefont {Rinn}, \citenamefont {Maass},\ and\ \citenamefont {Bouchaud}}]{rinn2001hopping}%
  \BibitemOpen
  \bibfield  {author} {\bibinfo {author} {\bibfnamefont {B.}~\bibnamefont {Rinn}}, \bibinfo {author} {\bibfnamefont {P.}~\bibnamefont {Maass}},\ and\ \bibinfo {author} {\bibfnamefont {J.-P.}\ \bibnamefont {Bouchaud}},\ }\bibfield  {title} {\bibinfo {title} {Hopping in the glass configuration space: subaging and generalized scaling laws},\ }\href@noop {} {\bibfield  {journal} {\bibinfo  {journal} {Physical Review B}\ }\textbf {\bibinfo {volume} {64}},\ \bibinfo {pages} {104417} (\bibinfo {year} {2001})}\BibitemShut {NoStop}%
\bibitem [{\citenamefont {Bertin}\ and\ \citenamefont {Bouchaud}(2003)}]{bertin2003subdiffusion}%
  \BibitemOpen
  \bibfield  {author} {\bibinfo {author} {\bibfnamefont {E.}~\bibnamefont {Bertin}}\ and\ \bibinfo {author} {\bibfnamefont {J.-P.}\ \bibnamefont {Bouchaud}},\ }\bibfield  {title} {\bibinfo {title} {Subdiffusion and localization in the one-dimensional trap model},\ }\href@noop {} {\bibfield  {journal} {\bibinfo  {journal} {Physical Review E}\ }\textbf {\bibinfo {volume} {67}},\ \bibinfo {pages} {026128} (\bibinfo {year} {2003})}\BibitemShut {NoStop}%
\bibitem [{\citenamefont {Burov}\ and\ \citenamefont {Barkai}(2007)}]{burov2007occupation}%
  \BibitemOpen
  \bibfield  {author} {\bibinfo {author} {\bibfnamefont {S.}~\bibnamefont {Burov}}\ and\ \bibinfo {author} {\bibfnamefont {E.}~\bibnamefont {Barkai}},\ }\bibfield  {title} {\bibinfo {title} {Occupation time statistics in the quenched trap model},\ }\href@noop {} {\bibfield  {journal} {\bibinfo  {journal} {Physical review letters}\ }\textbf {\bibinfo {volume} {98}},\ \bibinfo {pages} {250601} (\bibinfo {year} {2007})}\BibitemShut {NoStop}%
\bibitem [{\citenamefont {Hamdi}\ \emph {et~al.}(2024)\citenamefont {Hamdi}, \citenamefont {Burov},\ and\ \citenamefont {Barkai}}]{hamdi2024laplace}%
  \BibitemOpen
  \bibfield  {author} {\bibinfo {author} {\bibfnamefont {O.}~\bibnamefont {Hamdi}}, \bibinfo {author} {\bibfnamefont {S.}~\bibnamefont {Burov}},\ and\ \bibinfo {author} {\bibfnamefont {E.}~\bibnamefont {Barkai}},\ }\bibfield  {title} {\bibinfo {title} {Laplace's first law of errors applied to diffusive motion},\ }\href@noop {} {\bibfield  {journal} {\bibinfo  {journal} {arXiv preprint arXiv:2402.13733}\ } (\bibinfo {year} {2024})}\BibitemShut {NoStop}%
\bibitem [{\citenamefont {Akimoto}\ \emph {et~al.}(2018)\citenamefont {Akimoto}, \citenamefont {Barkai},\ and\ \citenamefont {Saito}}]{akimoto2018non}%
  \BibitemOpen
  \bibfield  {author} {\bibinfo {author} {\bibfnamefont {T.}~\bibnamefont {Akimoto}}, \bibinfo {author} {\bibfnamefont {E.}~\bibnamefont {Barkai}},\ and\ \bibinfo {author} {\bibfnamefont {K.}~\bibnamefont {Saito}},\ }\bibfield  {title} {\bibinfo {title} {Non-self-averaging behaviors and ergodicity in quenched trap models with finite system sizes},\ }\href@noop {} {\bibfield  {journal} {\bibinfo  {journal} {Physical Review E}\ }\textbf {\bibinfo {volume} {97}},\ \bibinfo {pages} {052143} (\bibinfo {year} {2018})}\BibitemShut {NoStop}%
\bibitem [{\citenamefont {Akimoto}\ and\ \citenamefont {Saito}(2020)}]{akimoto2020trace}%
  \BibitemOpen
  \bibfield  {author} {\bibinfo {author} {\bibfnamefont {T.}~\bibnamefont {Akimoto}}\ and\ \bibinfo {author} {\bibfnamefont {K.}~\bibnamefont {Saito}},\ }\bibfield  {title} {\bibinfo {title} {Trace of anomalous diffusion in a biased quenched trap model},\ }\href@noop {} {\bibfield  {journal} {\bibinfo  {journal} {Physical Review E}\ }\textbf {\bibinfo {volume} {101}},\ \bibinfo {pages} {042133} (\bibinfo {year} {2020})}\BibitemShut {NoStop}%
\bibitem [{\citenamefont {Machta}(1985)}]{machta1985random}%
  \BibitemOpen
  \bibfield  {author} {\bibinfo {author} {\bibfnamefont {J.}~\bibnamefont {Machta}},\ }\bibfield  {title} {\bibinfo {title} {Random walks on site disordered lattices},\ }\href@noop {} {\bibfield  {journal} {\bibinfo  {journal} {Journal of Physics A: Mathematical and General}\ }\textbf {\bibinfo {volume} {18}},\ \bibinfo {pages} {L531} (\bibinfo {year} {1985})}\BibitemShut {NoStop}%
\bibitem [{\citenamefont {Monthus}(2003)}]{monthus2003anomalous}%
  \BibitemOpen
  \bibfield  {author} {\bibinfo {author} {\bibfnamefont {C.}~\bibnamefont {Monthus}},\ }\bibfield  {title} {\bibinfo {title} {Anomalous diffusion, localization, aging, and subaging effects in trap models at very low temperature},\ }\href@noop {} {\bibfield  {journal} {\bibinfo  {journal} {Physical Review E}\ }\textbf {\bibinfo {volume} {68}},\ \bibinfo {pages} {036114} (\bibinfo {year} {2003})}\BibitemShut {NoStop}%
\bibitem [{\citenamefont {Ben~Arous}\ \emph {et~al.}(2006)\citenamefont {Ben~Arous}, \citenamefont {{\v{C}}ern{\`y}},\ and\ \citenamefont {Mountford}}]{ben2006aging}%
  \BibitemOpen
  \bibfield  {author} {\bibinfo {author} {\bibfnamefont {G.}~\bibnamefont {Ben~Arous}}, \bibinfo {author} {\bibfnamefont {J.}~\bibnamefont {{\v{C}}ern{\`y}}},\ and\ \bibinfo {author} {\bibfnamefont {T.}~\bibnamefont {Mountford}},\ }\bibfield  {title} {\bibinfo {title} {Aging in two-dimensional bouchaud's model},\ }\href@noop {} {\bibfield  {journal} {\bibinfo  {journal} {Probability theory and related fields}\ }\textbf {\bibinfo {volume} {134}},\ \bibinfo {pages} {1} (\bibinfo {year} {2006})}\BibitemShut {NoStop}%
\bibitem [{\citenamefont {Arous}\ and\ \citenamefont {{\v{C}}ern{\`y}}(2007)}]{arous2007scaling}%
  \BibitemOpen
  \bibfield  {author} {\bibinfo {author} {\bibfnamefont {G.~B.}\ \bibnamefont {Arous}}\ and\ \bibinfo {author} {\bibfnamefont {J.}~\bibnamefont {{\v{C}}ern{\`y}}},\ }\bibfield  {title} {\bibinfo {title} {Scaling limit for trap models on},\ }\href@noop {} {\bibfield  {journal} {\bibinfo  {journal} {The Annals of Probability}\ ,\ \bibinfo {pages} {2356}} (\bibinfo {year} {2007})}\BibitemShut {NoStop}%
\bibitem [{\citenamefont {Arous}\ \emph {et~al.}(2015)\citenamefont {Arous}, \citenamefont {Cabezas}, \citenamefont {Čern{\'y}},\ and\ \citenamefont {Royfman}}]{ben2015randomly}%
  \BibitemOpen
  \bibfield  {author} {\bibinfo {author} {\bibfnamefont {G.~B.}\ \bibnamefont {Arous}}, \bibinfo {author} {\bibfnamefont {M.}~\bibnamefont {Cabezas}}, \bibinfo {author} {\bibfnamefont {J.}~\bibnamefont {Čern{\'y}}},\ and\ \bibinfo {author} {\bibfnamefont {R.}~\bibnamefont {Royfman}},\ }\bibfield  {title} {\bibinfo {title} {{Randomly trapped random walks}},\ }\href {https://doi.org/10.1214/14-AOP939} {\bibfield  {journal} {\bibinfo  {journal} {The Annals of Probability}\ }\textbf {\bibinfo {volume} {43}},\ \bibinfo {pages} {2405 } (\bibinfo {year} {2015})}\BibitemShut {NoStop}%
\bibitem [{\citenamefont {{\v{C}}ern{\`y}}\ and\ \citenamefont {Wassmer}(2015)}]{vcerny2015randomly}%
  \BibitemOpen
  \bibfield  {author} {\bibinfo {author} {\bibfnamefont {J.}~\bibnamefont {{\v{C}}ern{\`y}}}\ and\ \bibinfo {author} {\bibfnamefont {T.}~\bibnamefont {Wassmer}},\ }\bibfield  {title} {\bibinfo {title} {Randomly trapped random walks on zd},\ }\href@noop {} {\bibfield  {journal} {\bibinfo  {journal} {Stochastic Processes and their Applications}\ }\textbf {\bibinfo {volume} {125}},\ \bibinfo {pages} {1032} (\bibinfo {year} {2015})}\BibitemShut {NoStop}%
\bibitem [{\citenamefont {Burov}(2017)}]{burov2017quenched}%
  \BibitemOpen
  \bibfield  {author} {\bibinfo {author} {\bibfnamefont {S.}~\bibnamefont {Burov}},\ }\bibfield  {title} {\bibinfo {title} {From quenched disorder to continuous time random walk},\ }\href@noop {} {\bibfield  {journal} {\bibinfo  {journal} {Physical Review E}\ }\textbf {\bibinfo {volume} {96}},\ \bibinfo {pages} {050103} (\bibinfo {year} {2017})}\BibitemShut {NoStop}%
\bibitem [{\citenamefont {Shafir}\ and\ \citenamefont {Burov}(2022)}]{shafir2022case}%
  \BibitemOpen
  \bibfield  {author} {\bibinfo {author} {\bibfnamefont {D.}~\bibnamefont {Shafir}}\ and\ \bibinfo {author} {\bibfnamefont {S.}~\bibnamefont {Burov}},\ }\bibfield  {title} {\bibinfo {title} {The case of the biased quenched trap model in two dimensions with diverging mean dwell times},\ }\href@noop {} {\bibfield  {journal} {\bibinfo  {journal} {Journal of Statistical Mechanics: Theory and Experiment}\ }\textbf {\bibinfo {volume} {2022}},\ \bibinfo {pages} {033301} (\bibinfo {year} {2022})}\BibitemShut {NoStop}%
\bibitem [{\citenamefont {Burov}(2020)}]{burov2020transient}%
  \BibitemOpen
  \bibfield  {author} {\bibinfo {author} {\bibfnamefont {S.}~\bibnamefont {Burov}},\ }\bibfield  {title} {\bibinfo {title} {The transient case of the quenched trap model},\ }\href@noop {} {\bibfield  {journal} {\bibinfo  {journal} {Journal of Statistical Mechanics: Theory and Experiment}\ }\textbf {\bibinfo {volume} {2020}},\ \bibinfo {pages} {073207} (\bibinfo {year} {2020})}\BibitemShut {NoStop}%
\bibitem [{\citenamefont {Burov}\ and\ \citenamefont {Barkai}(2012)}]{burov2012weak}%
  \BibitemOpen
  \bibfield  {author} {\bibinfo {author} {\bibfnamefont {S.}~\bibnamefont {Burov}}\ and\ \bibinfo {author} {\bibfnamefont {E.}~\bibnamefont {Barkai}},\ }\bibfield  {title} {\bibinfo {title} {Weak subordination breaking for the quenched trap model},\ }\href@noop {} {\bibfield  {journal} {\bibinfo  {journal} {Physical Review E}\ }\textbf {\bibinfo {volume} {86}},\ \bibinfo {pages} {041137} (\bibinfo {year} {2012})}\BibitemShut {NoStop}%
\bibitem [{\citenamefont {Burov}\ and\ \citenamefont {Barkai}(2011)}]{burov2011time}%
  \BibitemOpen
  \bibfield  {author} {\bibinfo {author} {\bibfnamefont {S.}~\bibnamefont {Burov}}\ and\ \bibinfo {author} {\bibfnamefont {E.}~\bibnamefont {Barkai}},\ }\bibfield  {title} {\bibinfo {title} {Time transformation for random walks in the quenched trap model},\ }\href@noop {} {\bibfield  {journal} {\bibinfo  {journal} {Physical Review Letters}\ }\textbf {\bibinfo {volume} {106}},\ \bibinfo {pages} {140602} (\bibinfo {year} {2011})}\BibitemShut {NoStop}%
\bibitem [{\citenamefont {Klafter}\ and\ \citenamefont {Sokolov}(2011)}]{klafter2011first}%
  \BibitemOpen
  \bibfield  {author} {\bibinfo {author} {\bibfnamefont {J.}~\bibnamefont {Klafter}}\ and\ \bibinfo {author} {\bibfnamefont {I.~M.}\ \bibnamefont {Sokolov}},\ }\href@noop {} {\emph {\bibinfo {title} {First steps in random walks: from tools to applications}}}\ (\bibinfo  {publisher} {OUP Oxford},\ \bibinfo {year} {2011})\BibitemShut {NoStop}%
\bibitem [{\citenamefont {Abramowitz}\ \emph {et~al.}(1988)\citenamefont {Abramowitz}, \citenamefont {Stegun},\ and\ \citenamefont {Romer}}]{abramowitz1972handbook}%
  \BibitemOpen
  \bibfield  {author} {\bibinfo {author} {\bibfnamefont {M.}~\bibnamefont {Abramowitz}}, \bibinfo {author} {\bibfnamefont {I.~A.}\ \bibnamefont {Stegun}},\ and\ \bibinfo {author} {\bibfnamefont {R.~H.}\ \bibnamefont {Romer}},\ }\href@noop {} {\emph {\bibinfo {title} {Handbook of mathematical functions with formulas, graphs, and mathematical tables}}}\ (\bibinfo  {publisher} {American Association of Physics Teachers},\ \bibinfo {year} {1988})\BibitemShut {NoStop}%
\bibitem [{\citenamefont {Weiss}\ and\ \citenamefont {Weiss}(1994)}]{weiss1994aspects}%
  \BibitemOpen
  \bibfield  {author} {\bibinfo {author} {\bibfnamefont {G.~H.}\ \bibnamefont {Weiss}}\ and\ \bibinfo {author} {\bibfnamefont {G.~H.}\ \bibnamefont {Weiss}},\ }\href@noop {} {\emph {\bibinfo {title} {Aspects and applications of the random walk}}}\ (\bibinfo  {publisher} {Elsevier Science \& Technology},\ \bibinfo {year} {1994})\BibitemShut {NoStop}%
\bibitem [{\citenamefont {Barkai}(2001)}]{barkai2001fractional}%
  \BibitemOpen
  \bibfield  {author} {\bibinfo {author} {\bibfnamefont {E.}~\bibnamefont {Barkai}},\ }\bibfield  {title} {\bibinfo {title} {Fractional fokker-planck equation, solution, and application},\ }\href@noop {} {\bibfield  {journal} {\bibinfo  {journal} {Physical Review E}\ }\textbf {\bibinfo {volume} {63}},\ \bibinfo {pages} {046118} (\bibinfo {year} {2001})}\BibitemShut {NoStop}%
\bibitem [{\citenamefont {Barkai}\ and\ \citenamefont {Fleurov}(1998)}]{Barkai1998}%
  \BibitemOpen
  \bibfield  {author} {\bibinfo {author} {\bibfnamefont {E.}~\bibnamefont {Barkai}}\ and\ \bibinfo {author} {\bibfnamefont {V.~N.}\ \bibnamefont {Fleurov}},\ }\bibfield  {title} {\bibinfo {title} {Generalized einstein relation: A stochastic modeling approach},\ }\href {https://doi.org/10.1103/PhysRevE.58.1296} {\bibfield  {journal} {\bibinfo  {journal} {Phys. Rev. E}\ }\textbf {\bibinfo {volume} {58}},\ \bibinfo {pages} {1296} (\bibinfo {year} {1998})}\BibitemShut {NoStop}%
\bibitem [{\citenamefont {Monthus}(2004)}]{monthus2004nonlinear}%
  \BibitemOpen
  \bibfield  {author} {\bibinfo {author} {\bibfnamefont {C.}~\bibnamefont {Monthus}},\ }\bibfield  {title} {\bibinfo {title} {Nonlinear response of the trap model in the aging regime: Exact results in the strong-disorder limit},\ }\href@noop {} {\bibfield  {journal} {\bibinfo  {journal} {Physical Review E}\ }\textbf {\bibinfo {volume} {69}},\ \bibinfo {pages} {026103} (\bibinfo {year} {2004})}\BibitemShut {NoStop}%
\bibitem [{\citenamefont {Schwarcz}\ and\ \citenamefont {Burov}(2022)}]{Schwarcz2022}%
  \BibitemOpen
  \bibfield  {author} {\bibinfo {author} {\bibfnamefont {D.}~\bibnamefont {Schwarcz}}\ and\ \bibinfo {author} {\bibfnamefont {S.}~\bibnamefont {Burov}},\ }\bibfield  {title} {\bibinfo {title} {Self-assembly of two-dimensional, amorphous materials on a liquid substrate},\ }\href {https://doi.org/10.1103/PhysRevE.105.L022601} {\bibfield  {journal} {\bibinfo  {journal} {Phys. Rev. E}\ }\textbf {\bibinfo {volume} {105}},\ \bibinfo {pages} {L022601} (\bibinfo {year} {2022})}\BibitemShut {NoStop}%
\bibitem [{\citenamefont {Vezzani}\ \emph {et~al.}(2019)\citenamefont {Vezzani}, \citenamefont {Barkai},\ and\ \citenamefont {Burioni}}]{Vezani2019}%
  \BibitemOpen
  \bibfield  {author} {\bibinfo {author} {\bibfnamefont {A.}~\bibnamefont {Vezzani}}, \bibinfo {author} {\bibfnamefont {E.}~\bibnamefont {Barkai}},\ and\ \bibinfo {author} {\bibfnamefont {R.}~\bibnamefont {Burioni}},\ }\bibfield  {title} {\bibinfo {title} {Single-big-jump principle in physical modeling},\ }\href@noop {} {\bibfield  {journal} {\bibinfo  {journal} {Phys. Rev. E}\ }\textbf {\bibinfo {volume} {100}},\ \bibinfo {pages} {012108} (\bibinfo {year} {2019})}\BibitemShut {NoStop}%
\bibitem [{\citenamefont {Vezzani}\ \emph {et~al.}(2020)\citenamefont {Vezzani}, \citenamefont {Barkai},\ and\ \citenamefont {Burioni}}]{Vezanni2020}%
  \BibitemOpen
  \bibfield  {author} {\bibinfo {author} {\bibfnamefont {A.}~\bibnamefont {Vezzani}}, \bibinfo {author} {\bibfnamefont {E.}~\bibnamefont {Barkai}},\ and\ \bibinfo {author} {\bibfnamefont {R.}~\bibnamefont {Burioni}},\ }\bibfield  {title} {\bibinfo {title} {Rare events in generalized lévy walks and the big jump principle},\ }\href@noop {} {\bibfield  {journal} {\bibinfo  {journal} {Sci. Rep.}\ }\textbf {\bibinfo {volume} {10}},\ \bibinfo {pages} {2732} (\bibinfo {year} {2020})}\BibitemShut {NoStop}%
\bibitem [{\citenamefont {Singh}\ and\ \citenamefont {Burov}(2023)}]{Singh2023}%
  \BibitemOpen
  \bibfield  {author} {\bibinfo {author} {\bibfnamefont {R.~K.}\ \bibnamefont {Singh}}\ and\ \bibinfo {author} {\bibfnamefont {S.}~\bibnamefont {Burov}},\ }\bibfield  {title} {\bibinfo {title} {Universal to nonuniversal transition of the statistics of rare events during the spread of random walks},\ }\href {https://doi.org/10.1103/PhysRevE.108.L052102} {\bibfield  {journal} {\bibinfo  {journal} {Phys. Rev. E}\ }\textbf {\bibinfo {volume} {108}},\ \bibinfo {pages} {L052102} (\bibinfo {year} {2023})}\BibitemShut {NoStop}%
\bibitem [{\citenamefont {H{\"o}ll}\ \emph {et~al.}(2023)\citenamefont {H{\"o}ll}, \citenamefont {Nissan}, \citenamefont {Berkowitz},\ and\ \citenamefont {Barkai}}]{holl2023controls}%
  \BibitemOpen
  \bibfield  {author} {\bibinfo {author} {\bibfnamefont {M.}~\bibnamefont {H{\"o}ll}}, \bibinfo {author} {\bibfnamefont {A.}~\bibnamefont {Nissan}}, \bibinfo {author} {\bibfnamefont {B.}~\bibnamefont {Berkowitz}},\ and\ \bibinfo {author} {\bibfnamefont {E.}~\bibnamefont {Barkai}},\ }\bibfield  {title} {\bibinfo {title} {Controls that expedite first-passage times in disordered systems},\ }\href@noop {} {\bibfield  {journal} {\bibinfo  {journal} {Physical Review E}\ }\textbf {\bibinfo {volume} {108}},\ \bibinfo {pages} {034124} (\bibinfo {year} {2023})}\BibitemShut {NoStop}%
\bibitem [{\citenamefont {Hughes}(1995)}]{hughes1995random}%
  \BibitemOpen
  \bibfield  {author} {\bibinfo {author} {\bibfnamefont {B.~D.}\ \bibnamefont {Hughes}},\ }\href@noop {} {\emph {\bibinfo {title} {Random walks and random environments: random walks}}},\ Vol.~\bibinfo {volume} {1}\ (\bibinfo  {publisher} {Oxford University Press},\ \bibinfo {year} {1995})\BibitemShut {NoStop}%
\bibitem [{\citenamefont {Montroll}\ and\ \citenamefont {Scher}(1973)}]{montroll1973random}%
  \BibitemOpen
  \bibfield  {author} {\bibinfo {author} {\bibfnamefont {E.~W.}\ \bibnamefont {Montroll}}\ and\ \bibinfo {author} {\bibfnamefont {H.}~\bibnamefont {Scher}},\ }\bibfield  {title} {\bibinfo {title} {Random walks on lattices. iv. continuous-time walks and influence of absorbing boundaries},\ }\href@noop {} {\bibfield  {journal} {\bibinfo  {journal} {Journal of Statistical Physics}\ }\textbf {\bibinfo {volume} {9}},\ \bibinfo {pages} {101} (\bibinfo {year} {1973})}\BibitemShut {NoStop}%
\bibitem [{\citenamefont {Gradshteyn}\ and\ \citenamefont {Ryzhik}(2014)}]{gradshteyn2014table}%
  \BibitemOpen
  \bibfield  {author} {\bibinfo {author} {\bibfnamefont {I.~S.}\ \bibnamefont {Gradshteyn}}\ and\ \bibinfo {author} {\bibfnamefont {I.~M.}\ \bibnamefont {Ryzhik}},\ }\href@noop {} {\emph {\bibinfo {title} {Table of integrals, series, and products}}}\ (\bibinfo  {publisher} {Academic press},\ \bibinfo {year} {2014})\BibitemShut {NoStop}%
\bibitem [{\citenamefont {Brummelhuis}\ and\ \citenamefont {Hilhorst}(1988)}]{brummelhuis1988single}%
  \BibitemOpen
  \bibfield  {author} {\bibinfo {author} {\bibfnamefont {M.}~\bibnamefont {Brummelhuis}}\ and\ \bibinfo {author} {\bibfnamefont {H.}~\bibnamefont {Hilhorst}},\ }\bibfield  {title} {\bibinfo {title} {Single-vacancy induced motion of a tracer particle in a two-dimensional lattice gas},\ }\href@noop {} {\bibfield  {journal} {\bibinfo  {journal} {Journal of statistical physics}\ }\textbf {\bibinfo {volume} {53}},\ \bibinfo {pages} {249} (\bibinfo {year} {1988})}\BibitemShut {NoStop}%
\bibitem [{\citenamefont {Riley}\ \emph {et~al.}(1999)\citenamefont {Riley}, \citenamefont {Hobson},\ and\ \citenamefont {Bence}}]{riley1999mathematical}%
  \BibitemOpen
  \bibfield  {author} {\bibinfo {author} {\bibfnamefont {K.~F.}\ \bibnamefont {Riley}}, \bibinfo {author} {\bibfnamefont {M.~P.}\ \bibnamefont {Hobson}},\ and\ \bibinfo {author} {\bibfnamefont {S.~J.}\ \bibnamefont {Bence}},\ }\href@noop {} {\bibinfo {title} {Mathematical methods for physics and engineering}} (\bibinfo {year} {1999})\BibitemShut {NoStop}%
\end{thebibliography}%

\end{document}